\documentclass[reprint,amsmath,amssymb,aps,pra,floatfix]{revtex4-2}

\usepackage{graphicx}
\usepackage{dcolumn}
\usepackage{bm}
\usepackage{braket}
\usepackage{amsmath}
\usepackage{chngcntr}
\DeclareMathOperator{\Tr}{Tr}

\begin{document}

\title{Optical Readout of Coherent Nuclear Spins Beyond the NV Center Electron T\textsubscript{1}}

\author{Johnathan Kuan\textsuperscript{1}}
\author{Gregory D. Fuchs\textsuperscript{2,3}}%
 \email{gdf9@cornell.edu}
\affiliation{%
\textsuperscript{1}Department of Physics, Cornell University, Ithaca, New York 14853, USA \\
\textsuperscript{2}School of Applied and Engineering Physics, Cornell University, Ithaca, New York 14853, USA \\
\textsuperscript{3}Kavli Institute at Cornell for Nanoscale Science, Ithaca, New York 14853, USA
}%

\begin{abstract}
The nitrogen-vacancy (NV) center is an emerging platform for constructing inertial sensors. Its native nitrogen spin can serve as a gyroscope using Ramsey interferometry protocols. The sensitivities of these nuclear-spin-based NV gyroscopes are limited by the phase coherence time of the nuclear spin, which is in turn limited by the $T_1$ of the NV center's electron spin. There is an active research effort to decouple the nuclear spin from the electron spin and other sources of decoherence to extend the phase coherence time past the NV center electron $T_1$. However,  this presents an associated challenge -- how to readout of the nuclear spin ensemble if the electronic spins have relaxed into a thermal state. We introduce an optical repump pulse to repolarize electron spins and investigate its effect on the readout of nuclear spins in this regime. We find for ensemble-based sensors, our readout protocol has around a six-fold improvement in sensitivity around the excited state anti-level crossing at 500 G compared to low fields (200 G), despite the presence of deleterious spin flip-flop dynamics in the excited state. 
\end{abstract}

\maketitle


\section{Introduction}

The negatively charged nitrogen-vacancy (NV) center in diamond is a versatile quantum sensor that has been used in applications such as in magnetometry \cite{Barry2020, Schloss2018, Taylor2008, Maze2008, Clevenson2015, Mamin2013}, electric field sensing \cite{Dolde2011, Michl2019, Barson2021}, and thermometry \cite{Toyli2012, Toyli2013, Fukami2019}. Sensing with the NV center is possible due to its coupling to external fields that cause shifts in its spin resonance frequency that can be detected with suitable pulse sequences and fluorescence contrast (i.e.  Ramsey interferometry). There have been recent demonstrations extending these techniques to construct rotation sensors (gyroscopes) using both the NV center's electron spin \cite{Wood2018} and native nitrogen nuclear spin as the sensor \cite{Soshenko2021, Jarmola2021}. In particular, the nuclear spin is of interest since it has a lower gyromagnetic ratio as compared to the electron spin, which makes it a more robust rotation sensor that is less sensitive to magnetic noise. These sensors promise to have better bias stability and sensing volumes than to conventional gyroscopes (e.g. microelectromechanical systems, ring lasers) due to the atom-like nature of the NV center. Despite the success of these demonstrations, the sensitivity of current NV center nuclear-spin-based gyroscopes is not competitive with conventional gyroscopes.
\begin{figure*}
    \centering
    \includegraphics[width=1\linewidth]{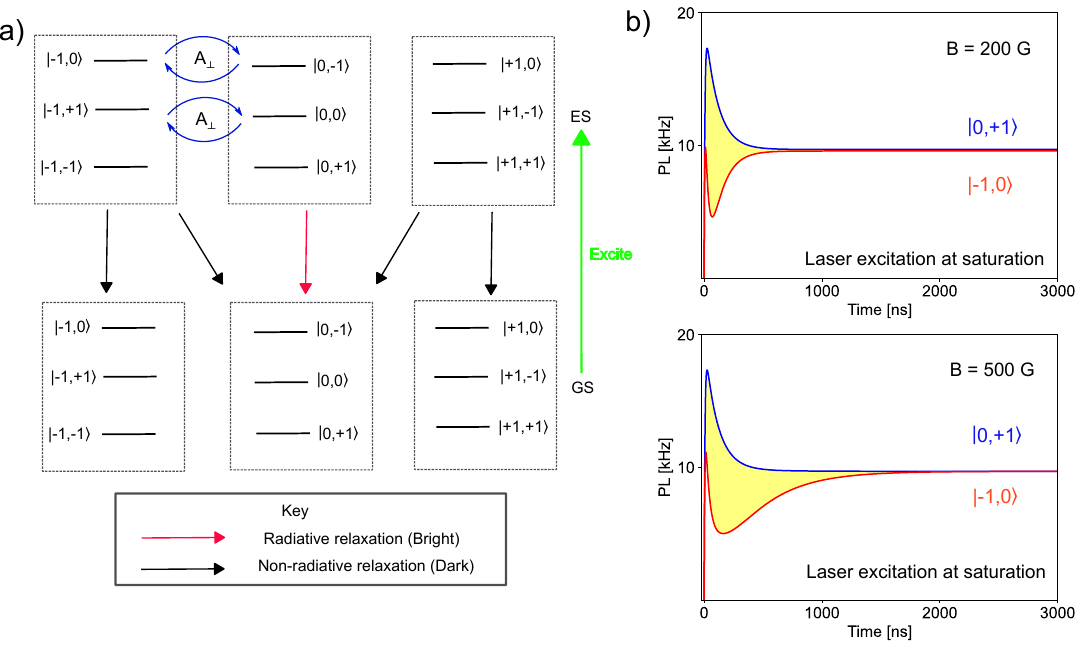}
    \caption{a) Optical cycling between the eigenstates of the ground state (GS) and excited state (ES). Radiative relaxation (red arrow)  only occurs for states in the ES $m_S=0$ manifold. Near the ESLAC (approximately 500~G), ES spin flip-flops (blue arrows) are mediated by the ES transverse hyperfine coupling $A_\perp$, which allows some electronic spins to move from the ES $m_S=0$ manifold to the ES $m_S=-1$ manifold. b) Lindblad calculation of the PL emission of $\ket{-1,0}$ and $\ket{0,+1}$ at 200~G (top) and 500~G (bottom) at optical saturation. The contrast (yellow) increases as the external field approaches the ESLAC due to the ES spin flip-flops.}
\end{figure*}

The sensitivity of current NV center nuclear-spin-based gyroscopes is fundamentally limited by the NV nuclear spin Ramsey coherence time $T_{2,n}^*$. Though the coherence time of an isolated nuclear spin can be on the order of seconds \cite{Ladd2005, Park2017, Shaham2022, Maurer2012, Steger2012}, the nuclear spin of the NV center is much shorter due to its hyperfine coupling to its neighboring electron spin. As the free precession time of a Ramsey sensing protocol approaches the electron $T_1$ of the NV center (on the order of 10 ms \cite{Jarmola2012}), the electron has a high chance of abruptly changing spin states (from $m_S=0$ to $m_S=\pm 1$). This $T_1$ process will change the nuclear spin precession via the electron spin state’s hyperfine coupling to the nuclear spin, which leads to the loss in nuclear spin coherence and limits the $T_{2,n}^*$ to the electron $T_1$.

There has been recent progress in extending the $T_{2,n}^*$ of the nuclear spin. Strategies involve decoupling the spin from noise sources including thermal and strain fluctuations in diamond, which allow sensing protocols to begin to approach the electron $T_1$ \cite{Wang2023, Wang2024}. Additionally, there are proposals to extend the nuclear spin coherence past the electron $T_1$ by driving strongly on the double quantum transition to average the hyperfine coupling to zero, which decouples the nuclear spin from the electron spin \cite{Chen2018}. 

In the context of these efforts, there is another fundamental problem -- reading out the nuclear spin for times exceeding the electron spin $T_1$. Conventionally, nuclear spin readout is done using a conditional mapping pulse (CNOT) \cite{Jiang2009, Neumann2010}. For times less than the electron $T_1$, the NV center is in the $m_S=0$ state. This allows a CNOT pulse (e.g. resonant with $\ket{0,0}$ to $\ket{-1,0}$) to map a single nuclear spin projection ($m_I = 0$) to an optically dark electron spin state ($m_S = -1$), creating nuclear spin-dependent photoluminescence (PL). However, the CNOT pulse is ineffective when we operate a nuclear sensor beyond the electron $T_1$ because the electron state becomes mixed over such long times, tending toward a thermal state. In this case, the system populations will be unchanged after the CNOT pulse because the nuclear spin is hyperfine coupled to each electron spin projection equally. Thus, there is a need to develop protocols to read out the nuclear spin in situations where it is used to sense with coherence times that exceed the electron $T_1$. Only after that is established, will it be be possible to observe electron-nuclear decoupling for sensing beyond $T_1$.

In this work, we test optical readout protocols of coherent nuclear spins with mixed thermal electron states. We use optical pumping to reset the electron spins back into the $m_S=0$ electron spin manifold before applying the CNOT pulse and study the efficiency of this process in retaining information stored in the nuclear spin. We focus on the nitrogen-14 isotope (spin-1) for our study because it is the most naturally abundant, however, the principles we discover are also applicable to nitrogen-15 NV centers.

\section{Readout Protocol for Nuclear Spins Beyond the Electron T\textsubscript{1}}
Our goal is to evaluate the impact of optically pumping before applying a CNOT pulse on the the shot-nose limited sensitivity of the sensor. For a nuclear spin-based gyroscope operating using a Ramsey protocol, this is given by
\begin{equation}
    \eta_{Ramsey}^{shot} \propto \frac{1}{V\sqrt{N T_{2,n}^*}}
\end{equation}
where $V$ is the Ramsey fringe visibility and $N$ is the ensemble size. The readout protocol will have an impact on the visibility of the Ramsey fringe due to the dynamics of optical pumping. 

The optical pumping dynamics are mainly due to the dynamics of the excited state (ES) \cite{Poggiali2017, Doherty2013, Duarte2021, Robledo2011, Gupta2016}. The ES Hamiltonian is
\begin{align}
    H_{ES} = D_{es}S_z^2 + \gamma_e B_z &+ PI_z^2 + \gamma_n B_z  + A_\parallel S_zI_z\\ &+ A_\perp(S_xI_x + S_y+ I_y) \nonumber
\end{align}
where $D_{es}$ is the zero field splitting in the excited state (1.42 GHz), $\gamma_e$ is the electron spin gyromagnetic ratio (2.8 MHz/G), $\gamma_n$ is the nuclear spin gyromagnetic ratio (0.3 kHz/G), $P$ is the quadrupolar moment of the nuclear spin ($-$4.85 MHz), and $A_\parallel$ and $A_\perp$ are the longitudinal and transverse hyperfine couplings ($-$43 MHz and $-$23 MHz respectively). The ground state (GS) Hamiltonian has a similar form
\begin{align}
    H_{GS} = D_{gs}S_z^2 + \gamma_e B_z &+ PI_z^2 + \gamma_n B_z  + C_\parallel S_zI_z\\ &+ C_\perp(S_xI_x + S_y+ I_y) \nonumber
\end{align}
where $D_{gs}$ is the zero field splitting in the ground state (2.87 GHz) and $C_\parallel$ and $C_\perp$ are the longitudinal and transverse hyperfine couplings (approximately 2.1 MHz). There are no significant dynamics in the ground state because the large $D_{gs}$ suppresses any spin mixing in its eigenstates at the fields we work at (200-800~G).

Near the excited state level anti-crossing (ESLAC) at approximately 500~G, the strong transverse hyperfine coupling in the excited state drives fast electron-nuclear spin flip-flops before the electronic orbital relaxation, which dominates the nuclear spin decoherence induced by optical pumping [Figure 1(a)]. These spin flip-flops are a useful resource for optically polarizing the nuclear spin but are undesirable from the standpoint of nuclear spin fidelity, because information obtained from sensing is erased \cite{Fischer2013, Fischer2013a, Jacques2009}. Therefore, there is a trade-off between generating electron polarization for more contrast and preserving nuclear spin fidelity.

The fast spin flip-flops present at fields around the ESLAC also provide a mechanism to enhance the sensitivity of the nuclear sensor. The spin flip-flops can prevent the electron from relaxing radiatively for multiple optical cycles for different spin states \cite{Steiner2010, Jarmola2020}. This causes a relative enhancement in contrast for certain states near the ESLAC, which may help improve the visibility of the Ramsey fringe from the sensing measurement [Figure 1(b)].
\begin{figure*}
    \centering
    \includegraphics[width=1\linewidth]{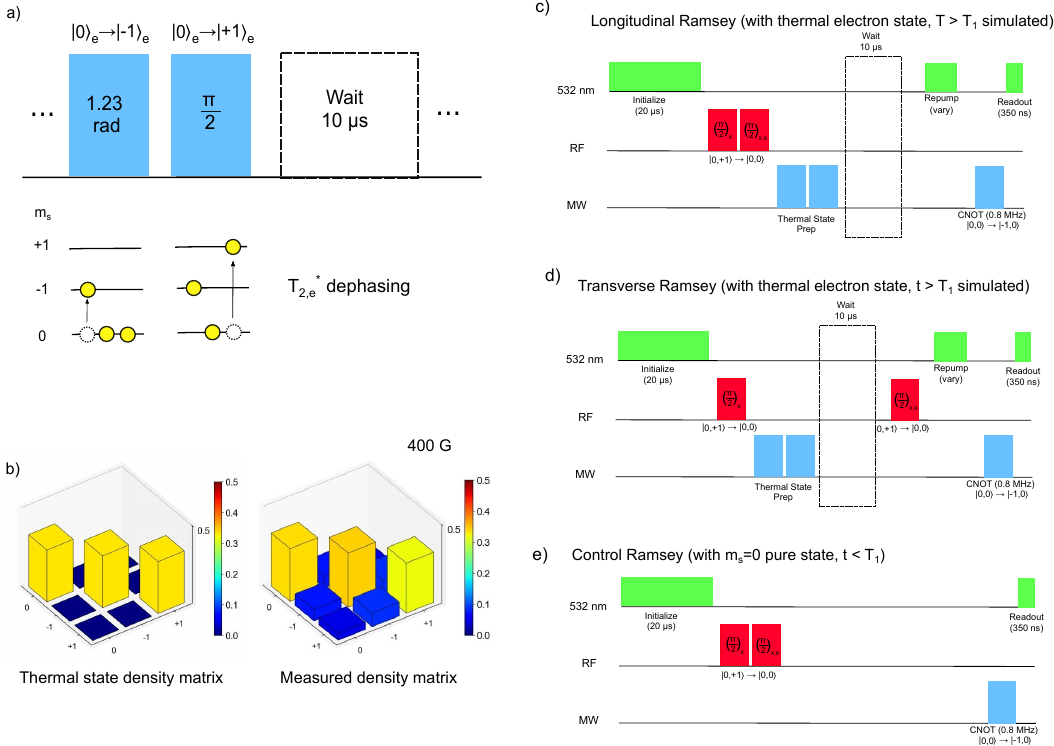}
    \caption{a) Artificial electron thermal state preparation. A pair of MW pulses is used to equalize the electron spin populations in all sublevels. A 10 $\mu$s wait, significantly greater than $T_{2,e}^*$, eliminates any coherences between sublevels. b) Quantum state tomography of the artificial thermal state at 400~G. The left shows the density matrix for a true thermal state (equal electronic spin populations and no coherences) and the right shows the measurement. The fidelity is approximately 98\%. c) Pulse sequence for the optical repump of the longitudinal component of the nuclear spin coupled to the artificial electron thermal state (simulating rotation sensing beyond the electron $T_1$). d) Pulse sequence for the optical repump of the transverse component of the nuclear spin coupled to the artificial electron thermal state (simulating rotation sensing beyond the electron $T_1$). e) Pulse sequence for control measurement. The nuclear spin is coupled to a pure $m_S=0$ state. This sequences measures the maximum visibility that could be theoretically obtained if the nuclear spin could be moved to the $m_S=0$ manifold without any loss of information.}
\end{figure*}

To evaluate the effect of these spin flip-flops on our readout protocol, we simulate a rotation sensing measurement beyond the electron $T_1$. To accurately simulate conditions of a real sensor, we use an ensemble of NV centers in an optical grade diamond from Element Six (approximately 150 defects/$\mu$m\textsuperscript{3}). For microwave control of the electron and nuclear spins, we use a platinum antenna fabricated on the diamond. PL is collected from the ensemble using a home-built confocal microscope. To minimize the effect of ensemble inhomogeneity on our measurements, we excite the NV ensemble at 1.6~mW, a factor of 10 below saturation.

 If the nuclear spin maintains its coherence past the electron $T_1$ due to some protocol that decouples the electron and nuclear spins, then the electron spin will be left in a mixed thermal state by spin-lattice relaxation. By the end of sensing protocol, the information stored in the nuclear spin is now coupled via the hyperfine interaction to this mixed electron thermal state. Therefore, a key component behind developing a readout protocol is to simulate the final state from sensing by coupling coherent nuclear spin states (i.e. pure states) to an artificial mixed thermal electron state. Explicitly, coupling a nuclear spin state $\rho_n$ to an artificial thermal electron state $\rho_e^{therm}$ means that we transform the state from $\ket{0}\bra{0}\otimes\rho_n $ to $\rho_e^{therm}\otimes\rho_n  $.
 
 To accomplish this, we first we prepare the nuclear spin state while the electron spin is in $m_S=0$. Then, the artificial thermal state of the electron spin is created using two microwave pulses and a waiting period [Figure 2(a)]. The first pulse, corresponding to a 1.23 radian rotation between the $m_S=0$ and $m_S=-1$ electron spin projections, moves $1/3$ of the electron spins to $m_S=-1$. A second $\pi/2$ pulse rotates half of the remaining $m_S=0$ into the $m_S=+1$ state. This pair of pulses equalizes the electron spin population in all three electron spin projections. To prevent disturbance of the nuclear spin, the pulses are significantly harder than $C_\perp$, at around 12 MHz. Finally, to eliminate coherences between the electron spin projections, we wait 10~$\mu$s after these pulses, much longer than the $T_{2,e}^*$ of the electron spin (around 600 ns). 

To assess the fidelity of our artificial thermal state, we perform quantum state tomography to measure the electron spin-1 density matrix \cite{Thew2002, Hofmann2004}. This requires measuring eight independent traceless spin observables (Gell-Mann matrices). Details of the state tomography can be found in Appendix B. After our measurement, we compare our artificial state to an ideal thermal electron state. Our procedure is robust, with a fidelity of approximately 98\% [Figure 2(b)].

After establishing a method to create artificial thermal states, we look at the effect of an optical repump pulse on the nuclear spin [Figure 3(c-d)]. The nuclear spin has two components: a longitudinal component (along $z$ on the Bloch sphere), and a transverse component (on the $xy$ plane of the Bloch sphere). For the longitudinal component, we look at the effect of the repump after the accumulated phase from sensing is mapped back to $z$ on the Bloch sphere for readout. For the transverse component, we look at the effect of the repump while the phase information remains on the $xy$ plane of the Bloch sphere. Looking at both these components gives us a complete picture of how the nuclear spin information is lost, which will guide us in developing the readout protocol.
\begin{figure*}
    \centering
    \includegraphics[width=1\linewidth]{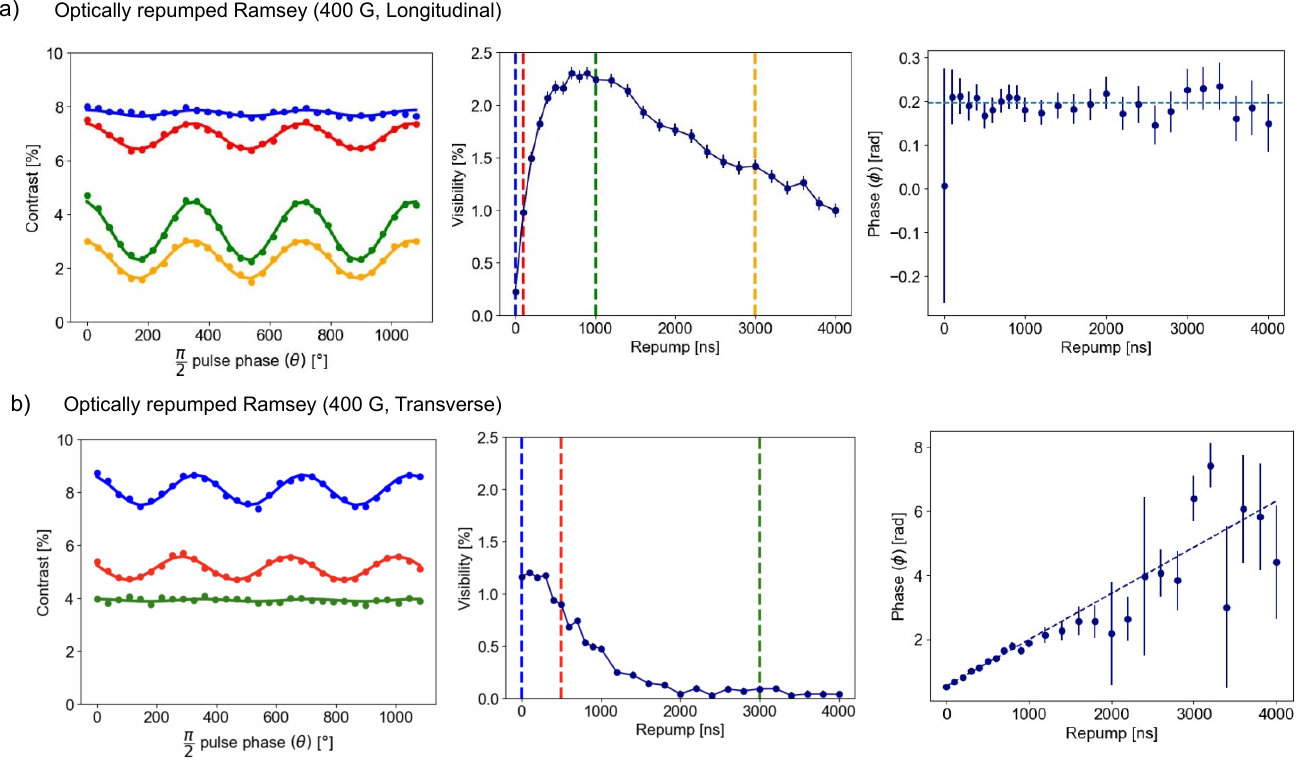}
    \caption{Representative measurements at 400~G of the visibility and initial phase $\phi$ of the Ramsey fringe for a) the longitudinal Ramsey sequence and b) the transverse Ramsey sequence. There are only significant increases in the visibility for the longitudinal Ramsey. The transverse Ramsey shows a linear dependence between the initial phase $\phi$ and the repump time.}
\end{figure*}

To simulate a rotation sensing measurement beyond the electron $T_1$, we use a Ramsey protocol where the coherent nuclear spin is hyperfine coupled to an artificial thermal state. For the longitudinal component of the nuclear spin, we first illuminate the NV centers for 20 $\mu$s with a 532 nm laser. This initializes the electron spin and nuclear spin to $\ket{0,+1}$. Next, a pair of $\pi/2$ pulses, resonant with the $\ket{0,+1}$ to $\ket{0,0}$ transition, is applied, with the second pulse phase shifted by $\theta$ relative to the first. After this, we perform our artificial thermal state preparation. We then apply the optical repump pulse to regain some electron polarization. To read out, we use a CNOT pulse that is resonant with the $\ket{0,0}$ to $\ket{-1,0}$ transition and illuminate with our 532 nm laser for 350 ns to collect the PL. For the transverse component of the nuclear spin, we perform a similar protocol, except the artificial thermal state is created between the two $\pi/2$ pulses. Finally, as a control we perform a conventional Ramsey protocol without the thermal state preparation [Figure 2(e)]. This control simulates the maximum signal we can theoretically achieve if we were able to move all the nuclear spins to the $m_S=0$ manifold without disturbing the nuclear spin.  We perform all these measurements (with and without the optical repump) at fields from 200-800 G.

To make the CNOT pulse selective relative to the hyperfine split nuclear states, we use a pulse that has a Rabi field of approximately 0.8 MHz ($\pi$ pulse length of 625 ns). This pulse, at worst, moves about 13\% of the population in $\ket{0,+1}$ to $\ket{-1,+1}$. It is difficult to make a highly selective CNOT pulse because its contrast is limited by ensemble inhomogeneity (i.e. $T_2^*\sim$400~ns). This can be caused by various factors, such as differences in the $B_1$ field experienced by individual NV centers in the ensemble. In addition, the contrast of the CNOT pulse will be affected by other sources of inhomogeneous broadening of the spin transition, such as dipolar interactions between the NV centers in the ensemble with other paramagnetic spins (i.e. other NV centers, P1 centers, and $^{13}$C nuclear spins). More information about the measured coherence time of our ensemble is presented in Appendix H.

This limitation imposed by inhomogeneous broadening is fundamental for sensing applications because the sensitivity is dependent on the ensemble size by $\sqrt{N}$. To maximize the sensitivity of our sensor, we want to work with a large number of nuclear spins and because nuclear spins have a small gyromagnetic ratio, we can work with a dense ensemble without significant inhomogeneous broadening. However, this increase in nuclear spin density comes with an increase in electron spin density from the NV centers associated with the nuclear spins and other paramgagnetic defects such as P1 centers. These electron spins have a gyromagnetic ratio that is orders of magnitude higher, so this $\sqrt{N}$ increase in sensitivity will come at the cost of inhomogenoeus broadening due to interactions between electron spins \cite{Acosta2009, Bauch2018, Bauch2020}. Thus, it is important to characterize the effect of these inhomogeneities on the our readout protocol in representative NV center ensembles (e.g. NV centers in optical grade diamond).

The Ramsey fringe generated by each measurement is fitted to extract the visibility and the initial phase. The PL of each data point is normalized to the PL emission from the $\ket{0,+1}$ state (Eq.~\ref{eq:contrast}) , which is independent of field. The Ramsey fringes, $S(\theta)$, are fitted to
\begin{equation}
S(\theta) = \frac{V}{2} \cos{(\theta + \phi)}+B
\end{equation}
where $V$ is the visibility (amplitude) of the Ramsey fringe, $B$ is the background, and $\theta$ is the relative angle between the $\pi/2$ pulses in radians. We measure $V$ over different repump times to track the impact the repump has on the readout sensitivity. Since $\phi$ relates to the acquired phase from sensing, it is important to see if it varies with our repump procedure.

\section{Results}
\begin{figure*}
    \centering
    \includegraphics[width=0.9\linewidth]{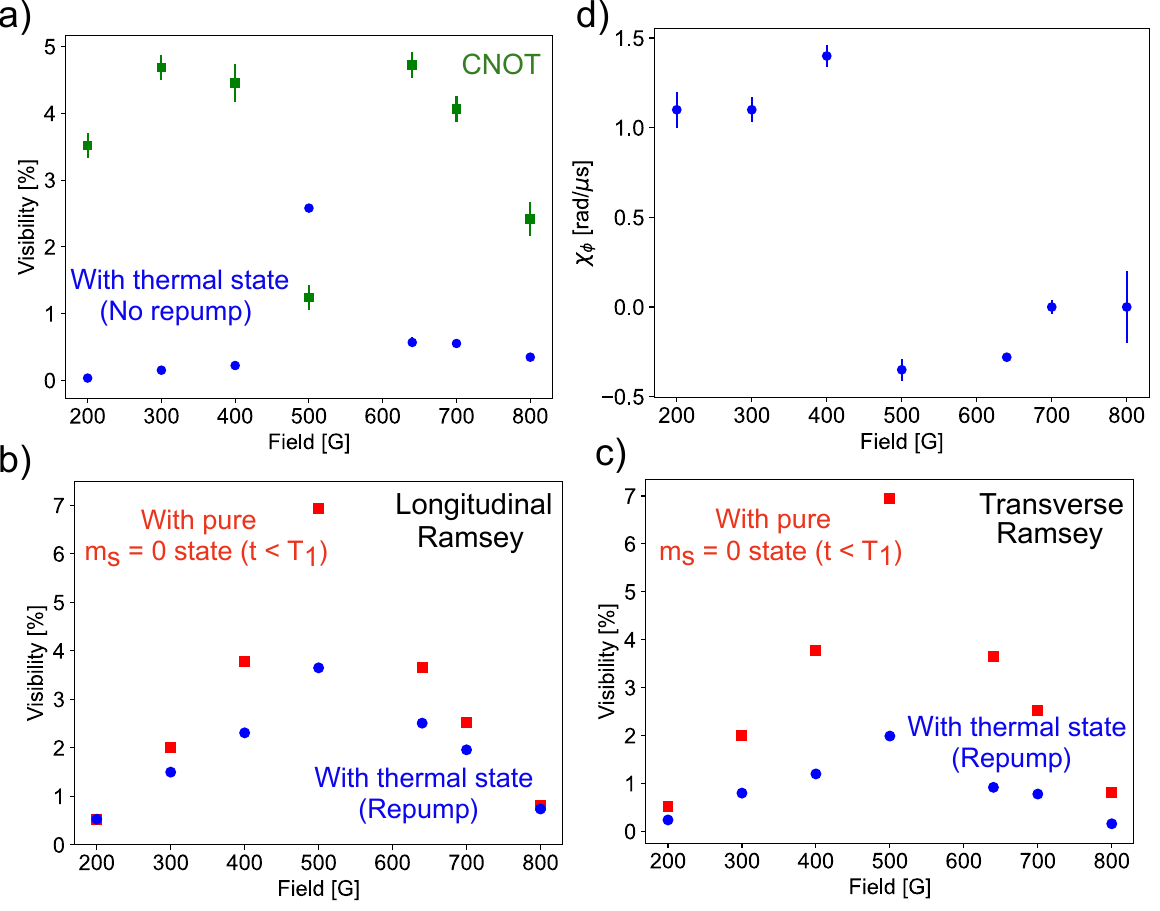}
    \caption{Summary of data collected from 200-800 G. a) Visibility of the CNOT readout when initialized to $\ket{0,0}$ (green). Maximum visibility of Ramsey sequences without a repump is shown in blue, which demonstrates the contribution of the contrast enhancement mechanism. b) Maximum recovered visibility for the longitudinal Ramsey (blue) versus the maximum possible visibility (red). The maximum occurs at 500 G, where the ES spin flip-flops occur at the highest rate. c)  Maximum recovered visibility for the transverse Ramsey (blue) versus the maximum possible visibility (red). The maximum also occurs 500 G, but is worse compared to the longitudinal Ramsey. d) Phase susceptibility $\chi_\phi$ for the transverse Ramsey from 200-800~G. The antisymmetric trend about 500 G suggests that the field dependence of the phase susceptibility is related to the ES dynamics.}
\end{figure*}

With the optical repump, the data shares the same qualitative features for all fields. To demonstrate these features, we show representative measurements at 400 G [Figure 3(a)]. For the longitudinal nuclear spin component, the visibility starts at a minimum with no repump. As the optical repump is applied for increasing duration, the visibility increases up to a maximum (approximately 1 $\mu$s for this field). The amount of recovery in the visibility depends on the field because it depends on competition between the gain in electron polarization with the loss in nuclear spin fidelity due to nuclear spin re-polarization. Past this point, the nuclear spin re-polarization process dominates, which leads to a decrease in visibility. 

We want to emphasize that we are exciting our NV ensemble with optical intensity that is a factor of 10 below saturation. However, our results should scale predictably with optical power (assuming we are sufficiently below saturation). Since the visibility depends only on the number of ES spin flip-flops that occur during the repump pulse, the maximum recovered visibility should be independent of excitation power. The repump time required to obtain this maximum visibility will change. We expect this time to be inversely proportional to the optical power, since increasing the laser excitation rate also increases the average number of spin flip-flops that occur, keeping the pulse length constant.

We also find that the initial phase from the fit is independent of pumping time. This is expected since the phase information from sensing is mapped back to $z$ prior to the repump. Semiclassically, the ES spin flip-flops modulate the longitudinal hyperfine coupling between the nuclear and electron spins because $m_S$ and $m_I$ will periodically change with time. This can generate an effective field for the nuclear spin to precess around during the optical cycling. However, since all the information is mapped back to $z$ before the repump, this precession will not have an effect on the nuclear spin populations, so the initial phase of the Ramsey fringe will not change.

For the transverse component of the nuclear spin [Figure 2(d)], there is no significant recovery in visibility [Figure 3(b)]. This is expected because the transverse component is more sensitive to the ES spin flip-flops. Because the nuclear spin is now on the transverse plane, the effective field created by the ES spin flip-flops will change the phase of the nuclear spin. In combination with stochastic relaxation back to the ground state (either through spontaneous emission or the dark intersystem crossing), this causes the phase of the transverse component on the Bloch sphere to undergo a random walk. Averaging over all these random walks over multiple optical cycles leads to a loss in the nuclear phase information, which translates to the loss in visibility of the observed Ramsey fringes. In comparison, the longitudinal component is insensitive to this precession and random walk, which enables the optical repump to recover some visibility of the Ramsey fringe.

When we look at the initial Ramsey phase $\phi$ extracted from the fit for the transverse component, we see that it increases linearly with repump time. Even though the phase undergoes a random walk, this stochastic process is biased in one direction, which appears as a slope in this plot. We will define the slope of this linear dependence as a phase susceptibility, $\chi_\phi$. When normalized to the laser excitation power, this susceptibility represents the average phase gain in one optical cycle from the transverse component’s random walk in the excited state.

Next, we use the Ramsey measurements without the optical repump to demonstrate the CNOT readout’s effectiveness [Figure 4(a)]. For fields excluding 500~G, the visibility is low because the is no electron polarization for the CNOT pulse to create nuclear spin-dependent PL. However, at 500~G, even though there is no electron polarization to generate contrast using the CNOT, we observe a high visibility of $\sim$2.5\%. In this case, the ES spin flip-flops cause nuclear spins in $m_I=0$ to relax through the dark ISC during optical readout more times on average than the spins in $m_I=+1$, creating contrast even where the electron is in a thermal mixed state. We also plot the visibility of the CNOT pulse, which is extracted from a Rabi oscillation of the electron spin from $\ket{0,0}$ to $\ket{-1,0}$. The visibility of the CNOT pulse dips to its minimum near the ESLAC, which is what we expect theoretically. The visibility from the CNOT pulse is the difference between the contrasts of $\ket{-1,0}$ and $\ket{0,0}$. Since the former reaches its minimum at the ESLAC and the latter its maximum, the overall difference will be lowest near the ESLAC (Figure A2). The CNOT visibility falls off as the field approaches 200 and 800 G due to poorer nuclear initialization to $\ket{0,+1}$.

For the optically repumped longitudinal component, we see that the maximum visibility is highest near the ESLAC, only a factor of two below the maximum theoretical visibility from the control measurement. The increase in visibility is a combination of two factors. The first factor, which dominates at fields away from the ESLAC, is the improved nuclear polarization as we approached the ESLAC. We extract the nuclear polarization from pulsed ODMR measurements which are discussed in Appendix F.  At low and high fields, the nuclear polarization into $m_I=+1$ is approximately 50\% and improves by approximately 35\% close to the ESLAC.  The second factor is the contrast enhancement due to the ES spin flip-flops, which primarily occurs near the ESLAC.  

The data indicate that the sensitivity gain from the PL contrast enhancement is comparable to the sensitivity loss from the nuclear polarization process. Compared to low fields (200 G), the maximum recovered visibility is a factor of 6 higher near the ESLAC (500 G). For the transverse component, the maximum recovered visibility is lower, a factor of four below the control measurement [Figure 4(c)].

 Our control measurements, with the electron spins polarized in the $m_S = 0$ spin state, represent the maximum visibility that our readout protocol can achieve in the limit where the electron $T_1$ is much longer than the sensing time. For a single NV center, we would expect that the visibility of the control measurement would be high away from the ESLAC and low near it. This is because all the spins in $\ket{0,0}$ are moved to $\ket{-1,0}$, so we expect the visibility to follow the same trend at the contrast of $\ket{-1,0}$ with field. However, the opposite trend appears in the data. This is another sign that the CNOT pulse is being limited by inhomogeneous broadening, which is a consequence of working with an ensemble of NV centers. The inhomogeneous broadening prevents us from moving all the spins from $\ket{0,0}$ to $\ket{-1,0}$. Therefore after the CNOT pulse is applied, the majority of spins still remain in $ket{0,0}$. This causes the contrast of this state, which reaches its maximum at the ESLAC, to dominate the trend observed in the visibility.
 
Finally, we also look at the phase susceptibility across 200-800 G [Figure 4(d)]. The plot of phase susceptibility with respect to field qualitatively appears to be anti-symmetric about 500 G. This provides strong evidence that the physical origin behind the phase susceptibility is related to the ESLAC dynamics. Considering the dependence of the phase susceptibility on the external field and a lack of recovery in the visibility with repump, optically repumping the transverse component is not an ideal way to recover nuclear spin information beyond the electron $T_1$.

\section{Discussion}
\begin{figure}
    \centering
    \includegraphics[width=1\linewidth]{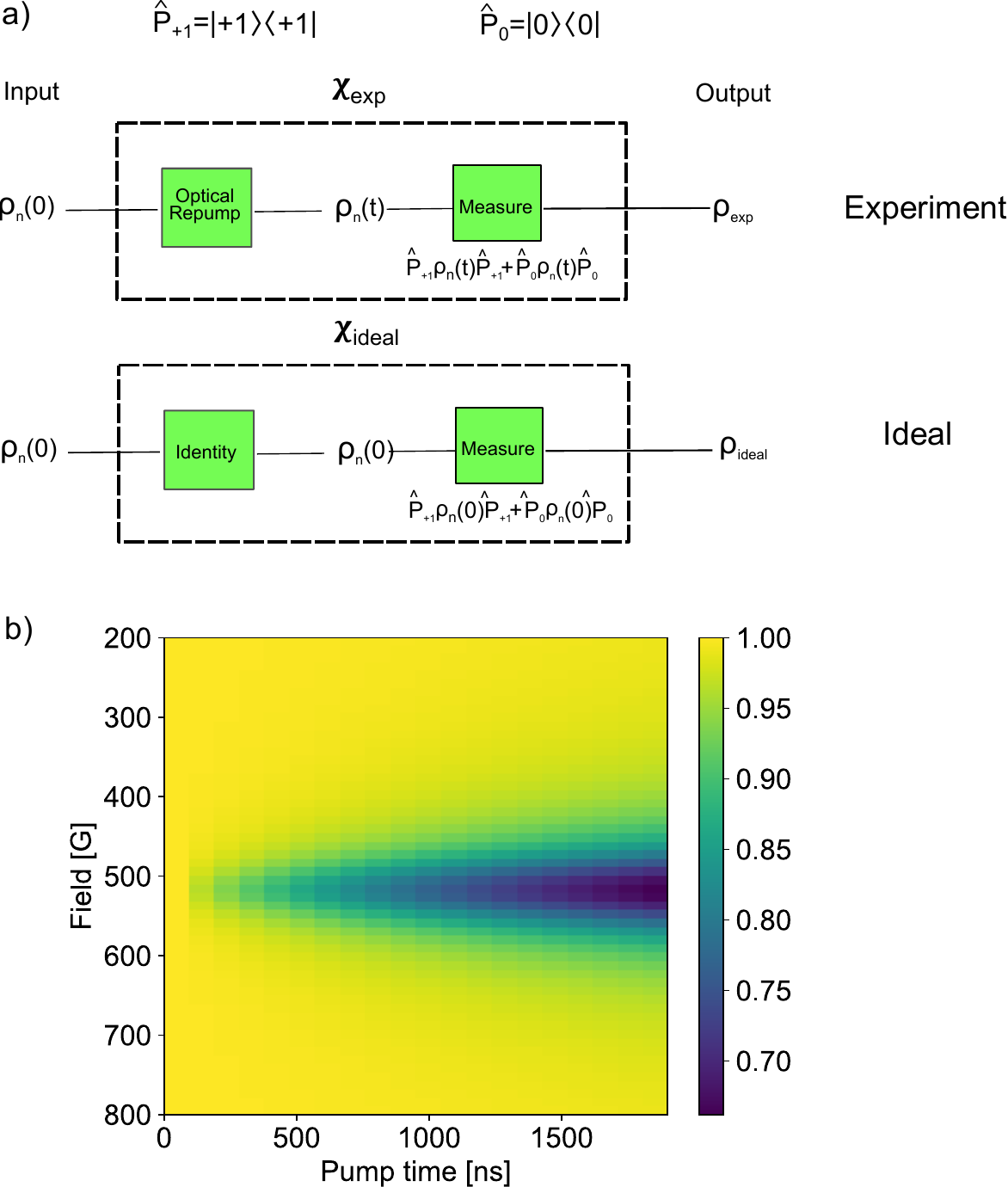}
    \caption{a) Outline of simulated quantum process tomography. The experimental process consists of the nuclear spin undergoing optical pumping, followed by measurement of the nuclear spin populations. The ideal process consists of the identity process on the nuclear spin (i.e. doing nothing), followed by measurement of the nuclear spin populations. b) Plot of the nuclear spin fidelity (process fidelity as defined in the main text) from 200-800 G for optical pumping up to 2 $\mu$s. The fidelity is the worst near the ESLAC (approximately 500~G) and for long pump times due to compounding the effect of many ES spin flip-flops.}
\end{figure}
From our measurements, we find that the maximum recovered visibility occurs at the ESLAC for both the longitudinal and transverse components of the nuclear spin. This suggests that the contrast enhancement that is generated from the ES spin flip-flops is comparable to the loss of phase information from the nuclear polarization process. This makes the 500~G ESLAC the best field to operate our readout protocols, despite the fact that it is also where the ES spin flip-flops occur at the highest rate. 

There are six different electron-nuclear spin states that are participating in the optical pumping ($\ket{0,+1}, \ket{\pm1,+1}, \ket{0,0}, \ket{\pm1,0}$) and it is not obvious which ones are generating the contrast enhancement that is compensating for the nuclear spin polarization process, as the PL emission from each state depends on the applied field. We want to understand this competition in a theoretical framework so that we can predict the performance of our readout protocol in other experimental settings (e.g. different optical powers, different nuclear spins) and improve its overall performance.

Because the experimental readout of the repumped transverse component of the nuclear spin is significantly worse than that of the repumped longitudinal component, we will focus on the physics of the latter for our readout protocol. We include a qualitative model for the observed dynamics of the transverse component in Appendix D.

We perform a 21 level Lindblad master equation calculation to understand how the contrast enhancement competes with the ES spin flip-flops.  The parameters and details for the Lindblad calculation can be found in Appendix A. The excitation rate for our calculation is chosen to match what is used in our measurements by comparing the incident laser power to the measured optical saturation power of our ensemble. The measurement used to extract the excitation rate is described in Appendix E.
\begin{figure*}
    \centering
    \includegraphics[width=1\linewidth]{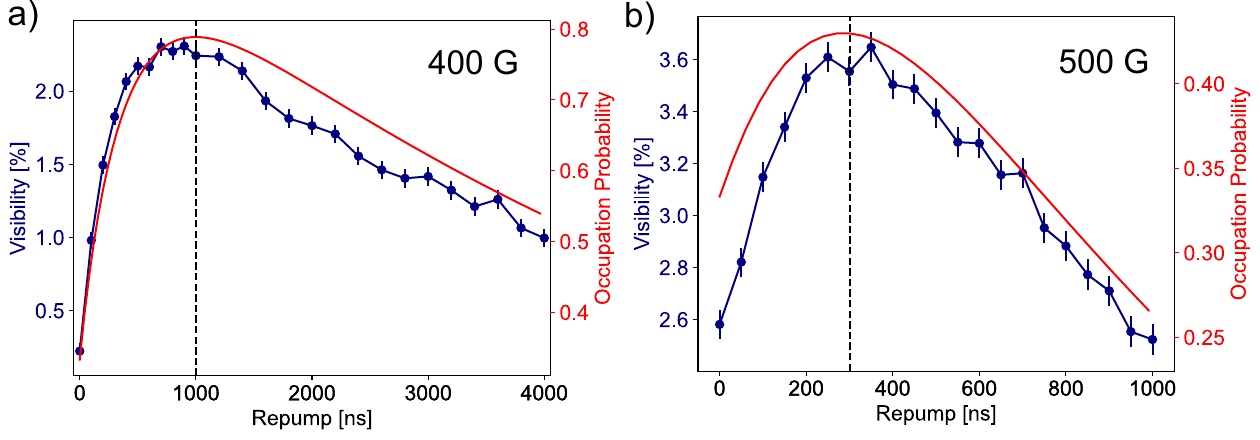}
    \caption{Comparison between the visibility of the longitudinal Ramsey over repump time and the calculated occupation in $\ket{0,0}$ for a) 400~G and b) 500~G. The maximum of the occupation from the calculation coincides with the maximum visibility, suggesting that $\ket{0,0}$ is the state that is determining the measured visibility.}
\end{figure*}

With the Lindblad model, we look at the effect of optical pumping on the nuclear spin fidelity at different magnetic fields. The nuclear polarization process eventually polarizes all the spins to $\ket{0,+1}$, decreasing the fidelity over time. We can quantify this fidelity by simulating quantum process tomography of the \{0, +1\} nuclear spin qubit \cite{Nielsen2012, Fuchs2012, Howard2006, O’Brien2004}. We treat the optical pumping of the NV center's electron spin and the ES spin flip-flops as an external source of dephasing. After calculating the final nuclear states after the optical repump, we calculate the process matrix, $\chi_{exp}$, corresponding to a measurement of the nuclear spin populations in $m_I=0$ and $m_I=+1$. We compare this to a process matrix, $\chi_{ideal}$, corresponding to an ideal measurement, where the nuclear spin populations are not affected by any source of dephasing [Figure 5(a)]. The fidelity is calculated from the process matrices as $F=\Tr[\sqrt{\chi_{exp}\chi_{ideal}}]$. 

Figure 5(b) shows the calculated nuclear fidelity for external fields from 200-800 G and for pump times up to 2~$\mu$s. The key feature is that the fidelity gets worse closer to the ESLAC and at increasing optical pump times. This is expected since, under these conditions, the ES spin flip-flops occur with high probability, which is the main source of decoherence. However, even at the ESLAC, there are still pump times (less than 500 ns) where the nuclear spin fidelity can be large (i.e. $F\geq 0.8$). It's in this regime where the contrast enhancement significantly compensates for the nuclear spin polarization process. Further details on our quantum process tomography calculation can be found in Appendix C.

Next, we examine the how the competing process, the field-dependent contrast enhancement, contributes the the visibility of the Ramsey fringe. The contrast enhancement of individual electron-nuclear spin states near the ESLAC depends significantly on the optical cycling of the electron spin, which varies for each state at different fields. We calculate the contrast for each state, relative to the PL emission from $\ket{0,+1}$, using our Lindblad model, where the NV center is initialized to a pure state of each electron-nuclear basis state. The contrasts for each nuclear spin state in each electron spin manifold can be found in Appendix A (Figure A2).

As discussed previously, a significant component to the observed trend in the visibility is the contrast enhancement of $\ket{0,0}$. Again, this is a consequence of the CNOT pulse being limited by inhomogeneous broadening. Another state that follows a similar trend in contrast as $\ket{0,0}$ is $\ket{0,-1}$ (Figure A2). However, we rule out $\ket{0,-1}$ a major contributors to the visibility since our nuclear qubit only involves $m_I=0$ and $m_I=+1$. While spins can enter $m_I=-1$ during the optical repump, our Lindblad calculations show that a negligible number of spins do this (Figure A3).  This suggests that the origin of the contrast enhancement, which is competing with the ES spin flip-flops, comes primarily from the spins in $\ket{0,0}$.

One observation is that the full width at half maximum (FWHM) of the repumped state's visibility is wider than the FWHM of the contrast of $\ket{0,0}$ observed in the Lindblad calculation. We attribute this to the change in nuclear polarization into $m_I = +1$, which varies over the entire field range ($50-80\%$). At fields away from the ESLAC, the visibility of the Ramsey fringe is dominated by two factors: the contrast of the CNOT readout and the polarization into $\ket{0,+1}$. At these fields, the contrast enhancement of $\ket{0,0}$ is negilible in comparison. As the field increases from 200~G to 300~G, the polarization drastically increases from approximately 50\% to approximately 80\%. This change in polarization allows for more spins to end up in $\ket{-1,0}$ after the CNOT readout. This leads to the apparent broadening of the repump state's visibility compared to the Lindblad calulation.

We can use the fact that the contrast enhancement is coming from $\ket{0,0}$ to predict the best optical repump pulse. The visibility is determined by spin population differences between two optically pumped initial states $\rho^e_{therm} \otimes \ket{+1}\bra{+1}$ and  $\rho^e_{therm} \otimes \ket{0}\bra{0}$. The effect of the $\ket{0,0}$ state’s contrast enhancement will largest when most of the spins are in $\ket{0,0}$ after optically pumping $\rho^e_{therm} \otimes \ket{0}\bra{0}$. To see this, we show, as examples, the calculated populations in $\ket{0,0}$ for this pumped state as a function of time for 400 G and 500 G. The observed maximum visibility coincides with the maximum population in $\ket{0,0}$ for both these fields [Figure 6(a-b)]. This provides a general rule for extending these results to other optical powers.

\section{Conclusion}
In summary, we investigate a readout protocol for coherent nuclear spins beyond the NV electron $T_1$, which involves repumping the mixed electron thermal state prior to a CNOT readout. We test our protocol on a simulated rotation sensing measurement beyond the NV electron $T_1$ and find that the sensitivity of our readout protocol is best near the ESLAC at 500 G, despite the fact that this is where the ES spin flip-flops occur the most.

There are several avenues to improve our readout protocol. One method is to engineer better microwave pulses for CNOT readout and to work with substrates and microwave structures that minimize sources of ensemble inhomogeneity. The $T_{2,e}^*$ coherence time for the NV center's electron spin is one measure of ensemble inhomogeneity. This requires engineering ensemble densities that balance the sensitivity loss from a shorter $T_{2,e}^*$ (limited by NV-NV dipolar interactions) with the $\sqrt{N}$ enhancement in sensitivity via the ensemble size. This also requires additional material engineering to minimize coupling to unwanted paramagnetic defects, which contribute to the inhomogeneous broadening of NV center's spin transitions. 

We can also improve the sensitivity of by changing our choice of CNOT readout. Using, a CNOT readout from $\ket{0,0}$ to $\ket{+1,0}$ would reduce the sensitivity loss since both states have contrast enhancement near the ESLAC. We did not use this specific CNOT rotation in our experiment, since this makes it difficult to disentangle the effects of the CNOT readout and contrast enhancement from each other.

Our technique for creating artificial thermal electron states can be applied to test readout protocols for other potential sold-state nuclear sensing platforms that are strongly hyperfine coupled to nearby electron spins, for example the Group-IV vacancy defects in diamond \cite{Harris2023, KuateDefo2021}. The most relevant application is extending our technique towards nitrogen-15 nuclear spins for NV centers. This isotope is of interest since, unlike the nitrogen-14 isotope, it has no quadrupolar moment. The quadrupolar moment of nitrogen-14 has been observed to have a temperature dependence, which causes laser heating of the diamond to be an additional source of decoherence \cite{Soshenko2020, Lourette2023}.

\section{Acknowledgements}
We thank Brendan McCullian, Anthony D’Addario, Ozan Erturk, and Sunil Bhave for helpful insights and discussions. This work was supported by the DOE Office of Science through Q-NEXT (National Quantum Information Science Research Centers) and the DARPA DRINQS program (Cooperative Agreement No. D18AC00024).  Device fabrication was done at the Cornell Nanoscale Facility, a member of the National Nanotechnology Coordinated Infrastructure (NNCI), which is supported by the National Science Foundation (Grant No. NNCI-2025233) and the Cornell Center for Materials Research Shared Facilities, which is supported through the NSF MRSEC program (Grant No. DMR-1719875).

\appendix
\counterwithin{figure}{section}
\section{Lindblad master equation calculations}
\begin{figure}
    \centering
    \includegraphics[width=1\linewidth]{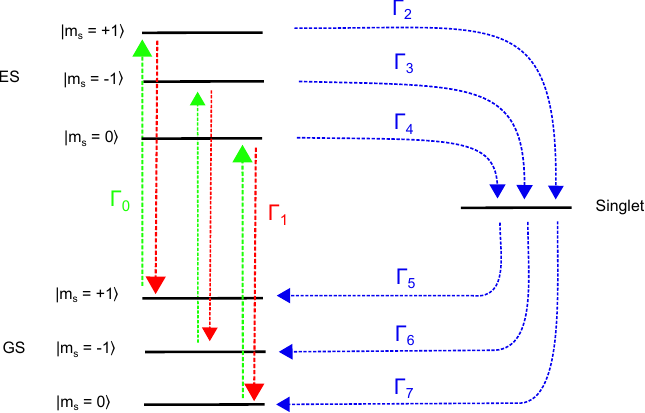}
    \caption{Transition rates used for the Lindblad model used in this work. Specific values are presented in Table 1. All transitions are assumed to preserve the nuclear spin quantum number ($m_I$).}
\end{figure}
To understand the effects of the NV center's optical cycling on the nuclear spin, we use the Lindblad master equation formalism \cite{Nielsen2012, Breuer2007}. We use the QuTiP python library for these calculations \cite{Qutip}. The Lindblad master equation is
\begin{equation}
    \frac{d\rho}{dt} = -i [H, \rho] + \sum_{i} \Gamma_i \left( L_i \rho L_i^\dagger - \frac{1}{2} \{L_i L_i^\dagger, \rho\}\right)
\end{equation}
where $L_i$ are the jump operators describing the non-unitary dynamics of the system and $\Gamma_i$ are the corresponding rates. Incoherent transitions between states $\ket{\psi_1}$ and $\ket{\psi_2}$ are described by the jump operator $\ket{\psi_1}\bra{\psi_2}$. Dephasing of individual states $\ket{\psi}$ is described by the jump operator $\ket{\psi}\bra{\psi}$. The rates used for our 21 level model (Figure A1) are presented in Table \ref{tab:table1}. We assume that all transitions are nuclear spin-preserving so there are only 8 unique rates. We also neglect $T_1$ and $T_2$ decoherence of the nuclear spin since these occur on much longer times scales than an optical cycle.

\begin{table}[b]
\caption{\label{tab:table1}%
Rates used for the 21 level Lindblad model (shown schematically in Figure A1). ES denotes the excited state and GS denotes the ground state. Rates are taken from Ref\cite{Gupta2016}.
}
\begin{ruledtabular}
\begin{tabular}{ccc}
\textrm{Transition$\backslash$Decoherence}&
\textrm{}&
\textrm{Rate (Linear Units)}\\
\colrule 
Excitation (GS to ES) [MHz]& $\Gamma_0$ & 6.74\vspace{0.1cm}\\ 
Spontaneous Emission [MHz] & $\Gamma_1$ & 67.4\vspace{0.1cm}\\
ES to singlet ($m_S = \pm1)$ [MHz] & $\Gamma_2$, $\Gamma_3$   & 91.6\vspace{0.1cm}\\
ES to singlet ($m_S = 0)$ [MHz] & $\Gamma_4$  & 9.9\vspace{0.1cm}\\
Singlet to GS ($m_S = \pm1)$ [MHz] & $\Gamma_5$ , $\Gamma_6$  & 1.06\vspace{0.1cm}\\
Singlet to GS ($m_S = 0)$ [MHz] & $\Gamma_7$  & 4.83\vspace{0.1cm}\\
Electron $T_1$ (GS) & $1/T_1^{GS}$ & $1/10$ ms\vspace{0.1cm}\\
Electron $T_2$ (GS) & $1/T_2^{GS}$  & $1/100$ $\mu$s\vspace{0.1cm}\\
Electron $T_1$ (ES) & $1/T_1^{ES}$  & $1/1$ ms\vspace{0.1cm}\\
Electron $T_2$ (ES) & $1/T_2^{ES}$  & $1/10$ ns\vspace{0.1cm}\\
\end{tabular}
\end{ruledtabular}
\end{table}

We use our Lindblad calculation to calculate the contrasts of all the electron-nuclear basis states from 200-800 G relative to $\ket{0,+1}$ (Figure A2). The contrast $C$ is calculated as
\begin{equation}
    C_{\ket{m_S,m_I}} = \dfrac{I_{\ket{0,+1}}-I_{\ket{m_S,m_I}}}{I_{\ket{0,+1}}}
\label{eq:contrast}
\end{equation}
where $PL_{\ket{0,+1}}$ is the PL emission from the NV center initialized to $\ket{0,+1}$ and $PL_{\ket{m_S,m_I}}$ is the PL emission from the NV center initialized to $\ket{m_S,m_I}$. The state $\ket{0,+1}$ is not affected by the ES spin flip-flops, which makes its PL emission independent of the applied external field. This allows us to use this state as a reference to compare the PL emission from the other electron-nuclear basis states. The majority of the states have an increased contrast near the ESLAC because the spin flip-flops in the excited state prevent the electron from relaxing radiatively. Other states have decreased contrast near the ESLAC (i.e $\ket{-1,0}$ and $\ket{-1,+1}$) because the spin flip-flops change the electron spin to $m_S=0$ in the excited state, which allows for radiative relaxation. The state $\ket{+1,+1}$ is unaffected by the spin flip-flops which causes its PL emission to be independent of field.

We neglect the effect of the spin flip-flop between the $\ket{0,-1}$ and $\ket{-1,0}$ when we calculate the nuclear spin fidelity with quantum process tomography. This is because a negligible number of nuclear spins enter the $m_I= -1$ spin manifold during our readout protocol, which means that the $\ket{0,0}$ and $\ket{-1,+1}$ spin flip-flop is the main driver of the excited state dynamics. To demonstrate this, we use the Lindblad model to track the total population in the $m_I = -1$ manifold for the optically pumped states $\rho_{therm}^e \otimes \ket{+1}\bra{+1}$ and $\rho_{therm}^e \otimes \ket{0}\bra{0}$ (Figure A3).

\begin{figure*}
    \centering
    \includegraphics[width=0.7\linewidth]{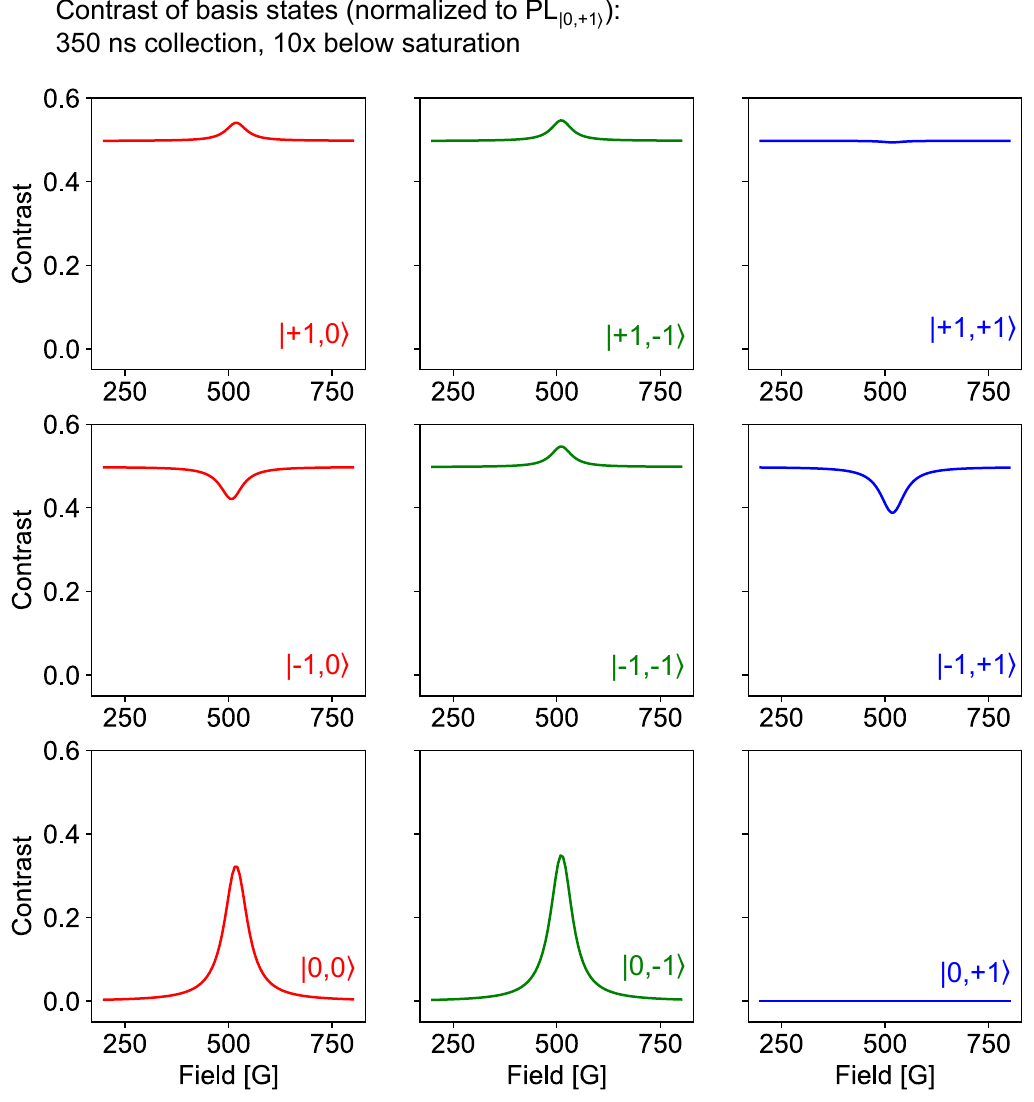}
    \caption{Calculated contrasts of all 9 basis states using the Lindblad model presented in Table I. The contrasts are calculated by normalizing to the PL emission from $\ket{0,+1}$ [Eq A2]. The collection time and power are chosen to match our measurements. Different states (e.g. $\ket{0,-1}$, $\ket{-1,+1}$) have an enhancement or loss in contrast at 500~G due to the ES spin flip-flops. The states $\ket{0,+1}$ and $\ket{+1,+1}$ are not affected by the ES spin flip-flops, making the contrasts independent of field.}
\end{figure*}

\begin{figure*}
    \centering
    \includegraphics[width=1\linewidth]{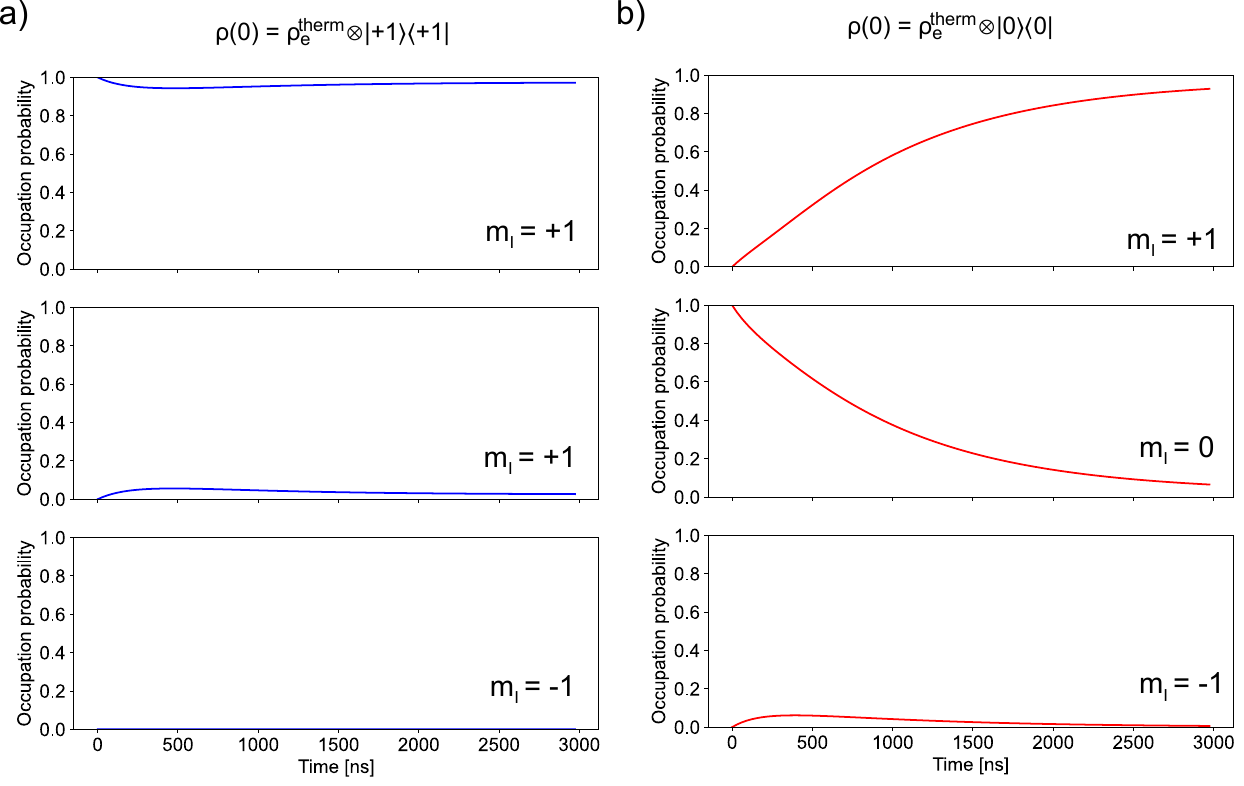}
    \caption{Lindblad calculation showing the occupations in the nuclear spin sublevels when the initial state a) $\rho_3^{therm} \otimes \ket{+1}\bra{+1}$ is optically pumped b) $\rho_3^{therm} \otimes \ket{0}
    \bra{0}$ is optically pumped. Both cases show negligible leakage into $m_I=-1$, which justifies neglecting the contribution of this sublevel.}
\end{figure*}

\section{Preparation and quantum state tomography of the artificial electron thermal state}
When we prepare our artificial thermal state, we ideally use a 1.23 rad pulse followed by a $\pi/2$ pulse. However, due to inhomogeneity of the $B_1$ field in the confocal volume, these pulses do not rotate the electron spin enough for some NV centers. To compensate for this, we measure the expectation value of $S_z$ from our prepared state when we use a 1.23 rad rotation and $\pi/2$ rotation. Then we adjust the rotations (i.e. over-rotate) until the measured expectation value is zero.

We use state tomography to check the fidelity of our artificial electron thermal state \cite{Thew2002, Hofmann2004}. To measure the spin-1 density matrix (qutrit), we parameterize the state using a basis of traceless Hermitian observables. This is given by
\begin{equation}
    \rho = \frac{1}{3}\mathbb{I}+\sum_{i=1}^8 \frac{1}{2}\langle\lambda_i\rangle\lambda_i
\end{equation}
where $\mathbb{I}$ is the identity operator, $\lambda_i$ are the Gell-mann matrices, and $\langle\lambda_i\rangle$ are the corresponding expectation values. The Gell-mann matrices are
\begin{equation}
\begin{split}
\lambda_1 &= \begin{pmatrix} 0 & 1 & 0 \\ 1 & 0 & 0 \\ 0 & 0 & 0\end{pmatrix}\quad
\lambda_2 = \begin{pmatrix} 0 & -i & 0 \\ i & 0 & 0 \\ 0 & 0 & 0\end{pmatrix}\\
\lambda_3 &= \begin{pmatrix} 1 & 0 & 0 \\ 0 & -1 & 0 \\ 0 & 0 & 0\end{pmatrix}\quad
\lambda_4 = \begin{pmatrix} 0 & 0 & 1 \\ 0 & 0 & 0 \\ 1 & 0 & 0\end{pmatrix}\\
\lambda_5 &= \begin{pmatrix} 0 & 0 & -i \\ 0 & 0 & 0 \\ i & 0 & 0\end{pmatrix}\quad
\lambda_6 = \begin{pmatrix} 0 & 0 & 0 \\ 0 & 0 & 1 \\ 0 & 1 & 0\end{pmatrix}\\
\lambda_7 &= \begin{pmatrix} 0 & 0 & 0 \\ 0 & 0 & -i \\ 0 & i & 0\end{pmatrix}\quad
\lambda_8 = \frac{1}{\sqrt{3}}\begin{pmatrix} 1 & 0 & 0 \\ 0 & 1 & 0 \\ 0 & 0 & -2\end{pmatrix}
\end{split}
\end{equation}
An intuitive way of thinking about these operators is as the Pauli operators for three different qubit subspaces (\{+1, 0\}, \{0, -1\}, and \{+1, -1\}). 

We measure the qutrit state in the interaction frame (i.e. the rotating frame of the NV center electron). To measure the expectation value of each spin operator, we apply microwave pulses resonant with the electron spin transitions. The set of applied pulses is equivalent to a set of unitary transformations that diagonalizes each $\lambda_i$. We read out the spin population of each spin-projection optically using a green laser pulse. For the $m_S = \pm 1$ populations, we use an adiabatic passage pulse to transfer the spins to $m_S = 0$ before the optical readout. The pulse sequences for all eight expectation values are shown in Figure B1(a-b). We measure our artificial electron thermal state when it is coupled to $\ket{m_I = +1}$.

To extract the spin populations in $m_S=+1$ and $m_S=-1$, we use calibration pulse sequences to measure the contrasts of $\ket{+1,+1}$ and $\ket{-1,+1}$ relative to $\ket{0, +1}$ (Figure B2). These calibration sequences are important because these contrasts are field-dependent, which complicates extracting the spin populations. We will denote $C_+$ and $C_-$ as the contrasts of $\ket{+1,+1}$ and $\ket{-1,+1}$ from the calibration sequences and $P_0$, $P_+$, and $P_-$ as the spin populations in $m_S=0$, $m_S=+1$, and $m_S=-1$ respectively.
\begin{figure*}
    \centering
    \includegraphics[width=0.8\linewidth]{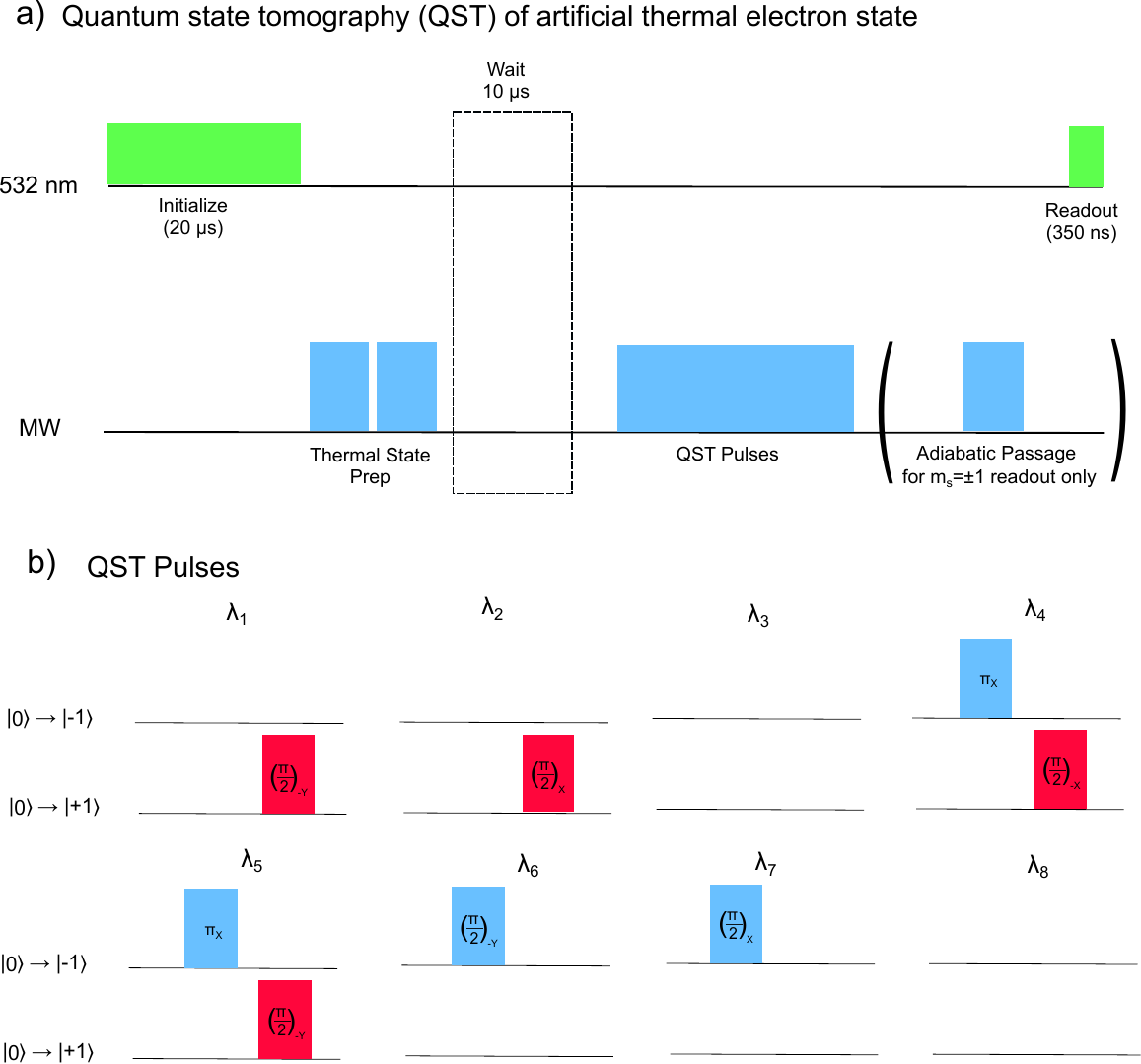}
    \caption{a) Quantum state tomography for checking artificial thermal state. After the thermal state is prepared, a set of QST pulses is used to find the expectation values of the spin operators $\lambda_i$ necessary for measuring the qutrit state. b) QST pulses used to diagonalize the spin operators before $S_z$ measurement.}
\end{figure*}
\begin{figure}
    \centering
    \includegraphics[width=1\linewidth]{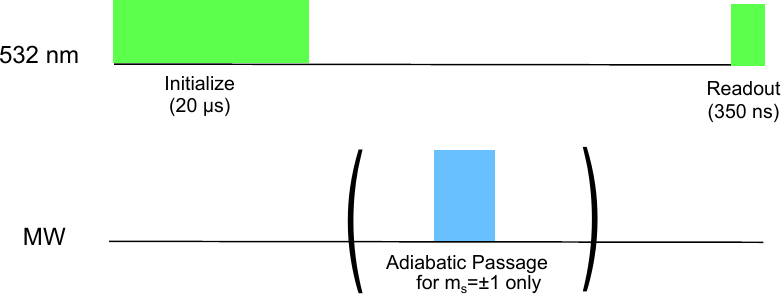}
    \caption{Calibration pulse sequences to measure PL emission from $m_S = 0, \pm 1$. The adiabatic passage pulses are applied after the 20 $\mu$s 532~nm laser excitation to prepare the electron spin to $m_S = \pm 1$.}
\end{figure}
When we measure the expectation value of each $\lambda_i$, we will measure three contrasts $R_0$, $R_+$, and $R_-$ corresponding to reading out the $m_S=0$, $m_S=+1$, and $m_S=-1$ populations. The contrasts from this measurement can be represented by the matrix equation
\begin{equation}
\begin{split}
R &= CP\\
\begin{pmatrix} R_0 \\ R_- \\R_+\end{pmatrix} &= \begin{pmatrix} 0 & C_- & C_+ \\  C_- & 0 & C_+ \\ C_+ & C_- &  0\end{pmatrix}\begin{pmatrix} P_0 \\ P_- \\P_+\end{pmatrix}
\end{split}
\end{equation}
The spin populations can then be determined by solving this equation.

After the qutrit is measured, the state may be unphysical (i.e. have negative eigenvalues and/or non-unit trace) due to experimental noise. To correct for this, we use a numerical estimation procedure, similar to maximum-likelihood estimation, which is outlined in Ref\cite{Howard2006}. To determine the measurement error for the individual elements of the qutrit state (from both the actual measurement and from the numerical estimation), we use statistical bootstrapping of our data. We find that the estimated measurement error from bootstrapping is approximately $1\% $. Additionally, there are systematic errors in the measurement that are caused by imperfect pulse fidelity and ensemble inhomogeneity. This systematic error is more difficult to directly quantify, but given that we use a wait much longer than $T_{2,e}^*$ in our thermal state preparation to eliminate any coherences, it should be on the order of the magnitude of the off-diagonal elements of the measured qutrit.

\section{Simulated quantum process tomography of the optical repump}
We use our 21 level Lindblad model to understand the effect of optical pumping of the NV center electron on the nuclear spin. The optical cycling of the electron and the excited state spin flip-flops are treated as an environmental source of decoherence for the nuclear spin. We compare the experimental process (simulated optical cycling followed by readout of the spin populations) to an ideal process (identity process followed by readout of the spin populations) using quantum process tomography [Figure 5(a)]. The tomography is used to calculate a process fidelity, which is referred to as the nuclear spin fidelity in the main text. This procedure allows us to quantify theoretically the degree to which nuclear spin information is lost due to the excited state spin flip-flops.

To simplify the calculation, we only work with final states projected in the \{0,+1\} subspace of the nuclear spin. As discussed previously (Appendix A), a negligible number of spins escape this subspace and leak into the $m_I=-1$ manifold, so this is a very good approximation. We calculate the process matrices for the simulated experimental process and the ideal process using the Pauli operator basis \{$\mathbb{I}, \sigma_x, \sigma_y, \sigma_z$\}. For the rest of this section, all the rows and columns of the process matrices will follow this ordering.

For the ideal process, the process matrix $\chi_{ideal}$ is given by
\begin{equation}
\chi_{ideal} = \begin{pmatrix} \frac{1}{2} & 0 & 0 & 0 \\  0 & 0 & 0 & 0 \\  0 & 0 & 0 & 0 \\ 0 & 0 & 0 &\frac{1}{2}\end{pmatrix}    
\end{equation}
or explicitly in the Kraus representation
\begin{equation}
    \varepsilon_{ideal}(\rho) = \frac{1}{2}\rho + \frac{1}{2}\sigma_z\rho\sigma_z
\end{equation}
where $\varepsilon_{ideal}(\rho)$ describes the evolution of the nuclear spin from the ideal process.

To find the process matrix for the simulated experimental process, we calculate the reduced density matrices of the nuclear spin by using our Lindblad model to evolve the initial states
\begin{equation} 
\begin{split}
\rho_1 = \rho_{e}^{therm}  \otimes \begin{pmatrix} 1 & 0 \\ 0 & 0\end{pmatrix}         \\
\rho_2 =  \rho_{e}^{therm}\otimes \begin{pmatrix} 0 & 1 \\ 0 & 0\end{pmatrix}    \\
\rho_3 =  \rho_{e}^{therm} \otimes \begin{pmatrix} 0 & 0 \\ 1 & 0\end{pmatrix}    \\
\rho_4 =  \rho_{e}^{therm} \otimes \begin{pmatrix} 0 & 0 \\ 0 & 1\end{pmatrix}    \\
\end{split}
\end{equation}
with
\begin{equation}
    \rho_{e}^{therm} = \begin{pmatrix} \frac{1}{3} & 0 & 0  \\  0 & \frac{1}{3} & 0  \\  0 & 0 & \frac{1}{3}  \end{pmatrix}
\end{equation}
For simplicity, we write the above initial states using a reduced Hilbert space consisting the ground state of the electron spin and the \{0, +1\} subspace of the nuclear spin. However, full 21-dimensional states are used as an initial input for the Lindblad calculations, where elements outside of this reduced Hilbert space are set to zero.

After the time evolution of the initial states, the system is allowed to relax for 1 $\mu$s before we calculate the reduced density matrices. These density matrices are calculated for different pumping times (up to 2 $\mu$s). Then for each pump time, we determine the process matrix $\chi_{exp}$. The process fidelity for each pump time is calculated as
\begin{equation}
    F = \Tr[\chi_{ideal}\chi_{exp}]
\end{equation}

This fidelity is a metric for how well the spin populations in each projection is preserved, where $F=1$ corresponds to the populations in both spin projections being perfectly preserved and $F=0.5$ corresponds to a complete loss of nuclear spin information due to the ES spin flip-flops. A simple way of thinking about the fidelity is as the square-root probability that the experimental process will give the same result as the ideal process. This procedure to calculate the nuclear spin fidelity is useful for optimizing our readout protocol for other optical powers and types of nuclear spins, which are not studied in this work.

\section{Qualitative model of the excited state phase susceptibility }
\begin{figure}
    \centering
    \includegraphics[width=1\linewidth]{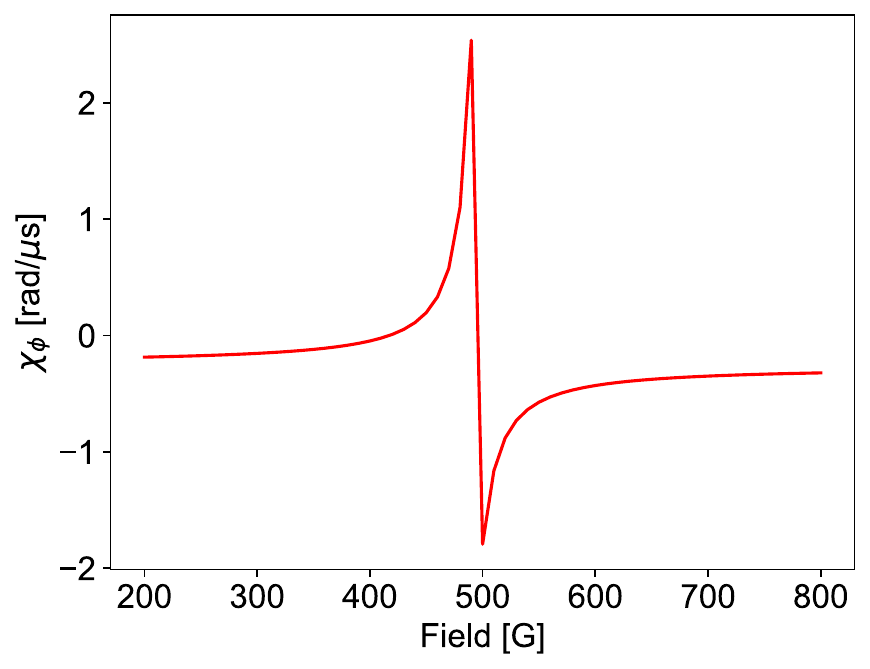}
    \caption{Plot of the calculated phase susceptibility from the four level ($\ket{0,0}$, $\ket{0,+1}$, $\ket{-1,0}$ , and $\ket{-1,+1}$) qualitative model, showing the antisymmetric dependence about the ESLAC. The excited state lifetime used for the calculation is 10~ns.}
\end{figure}
To understand the underlying physics of the phase susceptibility, we will focus on the effect that the excited state has on the nuclear spin and ignore the effect of the other states in the optical pumping cycle. The ES Hamiltonian is
\begin{align}
    H_{ES} = D_{es}S_z^2 + \gamma_e B_z &+ PI_z^2 + \gamma_n B_z  + A_\parallel S_zI_z\\ &+ A_\perp(S_xI_x + S_y+ I_y) \nonumber
\end{align}
where $D_{es}$ is the zero field splitting in the excited state (1.42 GHz), $\gamma_e$ is the electron spin gyromagnetic ratio (2.8 MHz/G), $\gamma_n$ is the nuclear spin gyromagnetic ratio (0.3 kHz/G), $P$ is the quadrupolar moment of the nuclear spin (-4.85 MHz), and $A_\parallel$ and $A_\perp$ are the longitudinal and transverse hyperfine couplings (-43 MHz and -23 MHz respectively). Because a negligible amount of spins leak into the $m_I=-1$ manifold, we will focus on the subspace consisting of the states $\ket{0,+1}$, $\ket{0,0}$, $\ket{-1,0}$, and $\ket{-1,+1}$. The Hamiltonian in matrix form is
\begin{equation}
H_{ES} = \begin{pmatrix} \omega_0 & 0 & 0 & 0 \\
0 & 0 & \frac{A_\perp}{2} & 0\\
0 & \frac{A_\perp}{2} & \omega_e + \omega_0 + A_\parallel & 0\\
0 & 0 & 0 & \omega_e
\end{pmatrix}  \\ 
\end{equation}
where
\begin{align}
    \omega_0 &= P + \gamma_n B_z \\
    \omega_e &= D_{ES} - \gamma_e B_z
\end{align}
are the Larmor frequencies for the nuclear and electron spin respectively. The transverse hyperfine coupling $A_\perp$ allows for coherent transitions between $\ket{0,0}$ and $\ket{-1,+1}$, which leads to the ES spin flip-flops.

To simplify the dynamics, we subtract out a global energy shift of $\dfrac{\omega}{2}$, where $\omega = \omega_0+\omega_e+ A_\parallel$
\begin{equation}
 H_{ES} =
 \frac{1}{2}\begin{pmatrix} 2\omega_0-\omega &  0 & 0 & 0\\
0 & -\omega  & A_\perp& 0\\
0 & A_\perp & \omega  & 0\\
0 & 0 & 0 & 2\omega_e - \omega
\end{pmatrix} 
\end{equation}
We then move to the rotating frame of the nuclear spin (RF1) with the unitary transformation $U_1$.
\begin{equation}
U_1 = \begin{pmatrix} e^{i\omega_0t/2} & 0 & 0 & 0 \\
0 & e^{-i\omega_0t/2}  & 0 & 0\\
0 & 0  & e^{i\omega_0t/2} & 0\\
0 & 0 & 0 & e^{-i\omega_0t/2}
\end{pmatrix}  \\ 
\end{equation}
which leads to the following effective Hamiltonian $H_1^{RF}$.
\begin{equation}
H_1^{RF} = \frac{1}{2}\begin{pmatrix} -\omega_e- A_\parallel & 0 & 0 & 0 \\
0 & -\omega_e - A_\parallel  & A_\perp e^{-i\omega_0t} & 0\\
0 & A_\perp e^{i\omega_0t} & \omega_e + A_\parallel  & 0\\
0 & 0 & 0 & \omega_e - A_\parallel
\end{pmatrix}  \\ 
\end{equation}
In the rotating frame of the nuclear spin, the hyperfine coupling appears as a time-varying field which drives Rabi oscillations (ES spin flip-flops). We will calculate the phase susceptibility in this frame, since this is what we measure experimentally.

To eliminate the time dependence in the $\ket{0,0}$ and $\ket{-1,+1}$ subspace, we apply another unitary transformation $U_2$.
\begin{equation}
U_2 = \begin{pmatrix} 1 & 0 & 0 & 0 \\
0 & e^{i\omega_0t/2}  & 0 & 0\\
0 & 0  & e^{-i\omega_0t/2} & 0\\
0 & 0 & 0 & 1
\end{pmatrix}  \\ 
\end{equation}
This gives a new effective Hamiltonian $H_2^{RF}$.
\begin{equation}
H_2^{RF} = \frac{1}{2}\begin{pmatrix} -\omega_e - A_\parallel & 0 & 0 & 0 \\
0 & -\omega & A_\perp & 0\\
0 & A_\perp & \omega  & 0\\
0 & 0 & 0 & \omega_e - A_\parallel
\end{pmatrix}  \\ 
\end{equation}
The time-evolution operator for the nuclear spin in the excited state is obtained by exponentiation.
\begin{equation}
\exp{\left(-iH_2^{RF}t\right)}=\begin{pmatrix} e^{i(\omega_e+A_\parallel) t/2} & 0 & 0 & 0 \\
0 & \alpha & -i\beta & 0\\
0 & -i\beta & \alpha^* & 0\\
0 & 0 & 0 & e^{-i(\omega_e-A_\parallel) t/2} \end{pmatrix}
\end{equation}
where
\begin{align}
\alpha &= \cos\left(\frac{\Omega t}{2}\right) + i \frac{\omega}{\Omega}\sin\left(\frac{\Omega t}{2}\right)\\
\beta &= \frac{A_\perp}{\Omega}\sin\left(\frac{\Omega t}{2}\right)\\
\Omega &= \sqrt{A_\perp^2 + \omega^2}
\end{align}
If we go back to the original rotating frame (RF1), then the time evolution operator is
\begin{equation}
U(t) = U_2^\dagger\exp\left(-iH_2^{RF}t\right)U_2
\end{equation}
In this frame (RF1), the initial state is
\begin{equation}
\begin{split}
    \rho_{e, n}^{therm}(0) &= \rho_e^{therm}(0) \otimes \rho_n(0)\\
    &=\left[ \frac{1}{2}\begin{pmatrix}1 & 0 \\ 0 & 1\end{pmatrix}\right] \otimes \left[ \frac{1}{2}\begin{pmatrix}   1 & e^{-i\Phi_0} \\ e^{i\Phi_0} & 1\end{pmatrix}\right]
\end{split}
\end{equation}
where the nuclear spin in on the equator of its Bloch sphere at some angle $\Phi_0$ relative to $x$ and the electron spin is in a thermal mixed state. We then apply $U(t)$ to our initial state and look at the matrix element $\braket{0,+1|\rho_{e, n}^{therm}(t)|0,0}$.
\begin{equation}
 \braket{0,+1|\rho_{e, n}^{therm}(t)|0,0} = \alpha^*\exp\left[-i(\Phi_0-\frac{\omega_e+ A_\parallel}{2} t)\right]   
\end{equation}

The complex phase $\Phi(t)$ of $\alpha^*$ is determined by the relation
\begin{equation}
\tan\Phi(t) = \dfrac{\omega}{\Omega}\tan\left (\frac{\Omega t}{2}\right)
\end{equation}
so the total phase $\phi$ in the matrix element is
\begin{equation}
\phi = \Phi_0+\Phi(t)-\frac{\omega_e + A_\parallel}{2}t
\end{equation}
The time derivative of $\Phi(t)$ is found through implicit differentiation and is given by
\begin{equation}
\frac{\partial\Phi(t)}{\partial t} =\dfrac{1}{2} \frac{\Omega^2 \omega}{\Omega^2 \cos^2(\frac{\Omega t}{2}) + \omega^2 \sin^2(\frac{\Omega t}{2})}
\end{equation}
Finally, the phase precession velocity is
\begin{equation}
\frac{\partial\phi}{\partial t} =\dfrac{1}{2}\left[ \frac{\Omega^2 \omega}{\Omega^2 \cos^2(\frac{\Omega t}{2}) + \omega^2 \sin^2(\frac{\Omega t}{2})} - \omega_e - A_\parallel\right ]
\end{equation}

To get the average phase, $\langle \Delta \phi \rangle$, gained in an optical cycle, we assume an average ES lifetime, $T$, of 10 ns.
\begin{equation}
    \langle \Delta \phi \rangle = \int_{0}^{\infty}\int_{0}^{t}\frac{1}{T}e^{-\frac{t}{T}}\frac{\partial\phi}{\partial t'} \, dt'\,  dt
\end{equation}
The phase susceptibility $\chi_\phi$ is given by
\begin{equation}
    \chi_\phi = \frac{ \langle \Delta \phi \rangle}{T}
\end{equation}
We show the result of this qualitative model in Figure D1, which captures the antisymmetric behavior of the phase susceptibility about the ESLAC. 

We want to emphasize that this is only a basic qualitative model since we have not included the complete Lindblad dynamics, which includes transitions involving the other 17 available states. For example, the electron polarization process will also modulate the hyperfine coupling in the ground state since it moves spins in $m_S=\pm1$ to $m_S=0$. This creates an additional effective field which will contribute to the phase precession of the transverse component of the nuclear spin. Since the electron polarization process is essentially independent of the external field, this process would appear as an constant offset in the susceptibility. In addition, we ignored the effect of the $\ket{0,-1}$ to $\ket{-1,0}$ spin flip-flop, which will add contributions to the susceptibility. Despite these limitations, this qualitative model provides a general picture of the physics behind the field-dependent phase susceptibility.

An interesting observation from this model is that as long as the coherences in the electron spin are small (i.e the electron spin remains sufficiently mixed during the optical pumping), the phase precession velocity is independent of the electron spin polarization. We can see this by replacing the initial state for the electron spin with a mixed state of any polarization. This means that that this model predicts that the phase susceptibility is independent of time, which is what we observe in our measurements [Figure 3(b)].

\section{Measurement of experimental laser excitation rate}
\begin{figure}
    \centering
    \includegraphics[width=1\linewidth]{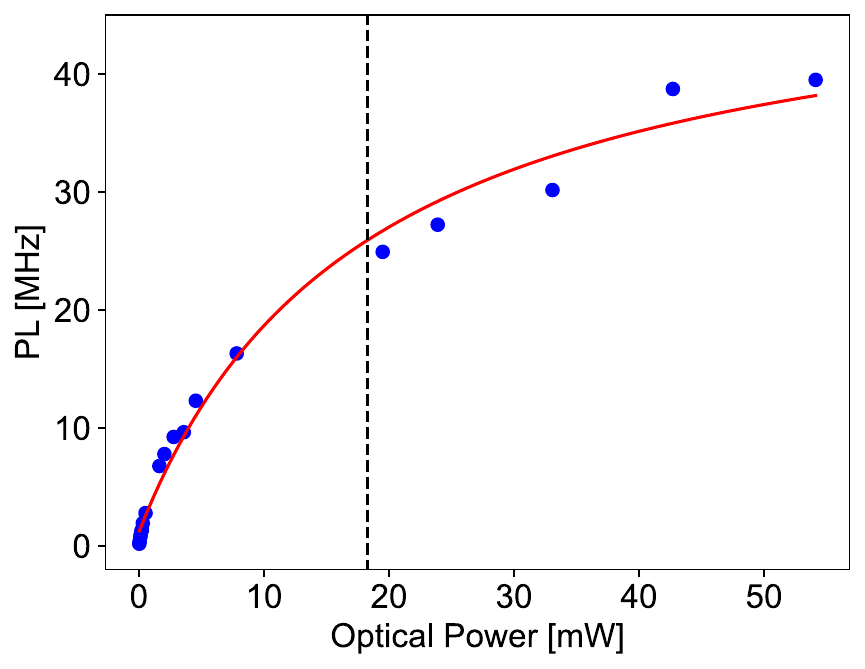}
    \caption{Plot of the NV center ensemble photoluminescence versus excitation power. The black dotted line indicates the saturation power extracted from the fit [Eq E1].}
\end{figure}
A key parameter for our Lindblad calculations is the laser excitation rate of the NV center electron spin from the ground state to the excited state. To determine this parameter we measure the optical saturation of our NV ensemble and fit our measurement to
\begin{equation}
    I(P) = \frac{I_0}{1 + P/P_{sat}}+B
\end{equation}
where $I_0$ corresponds to the saturation PL intensity, B is the background intensity, and $P_{sat}$ is the saturation power \cite{McCullian2022}. The saturation curve is shown in Figure E1. The extracted saturation power from the fit is $18 \pm 3$~mW. At saturation, the laser excitation rate is comparable to the spontaneous emission rate (67.4~MHz). Therefore we estimate the experimental excitation rate by scaling the spontaneous emission rate by the ratio of the excitation power we use (1.6~mW) to the saturation power.

\section{Nuclear spin polarization from 200-800~G}
\begin{figure}
    \centering
    \includegraphics[width=1\linewidth]{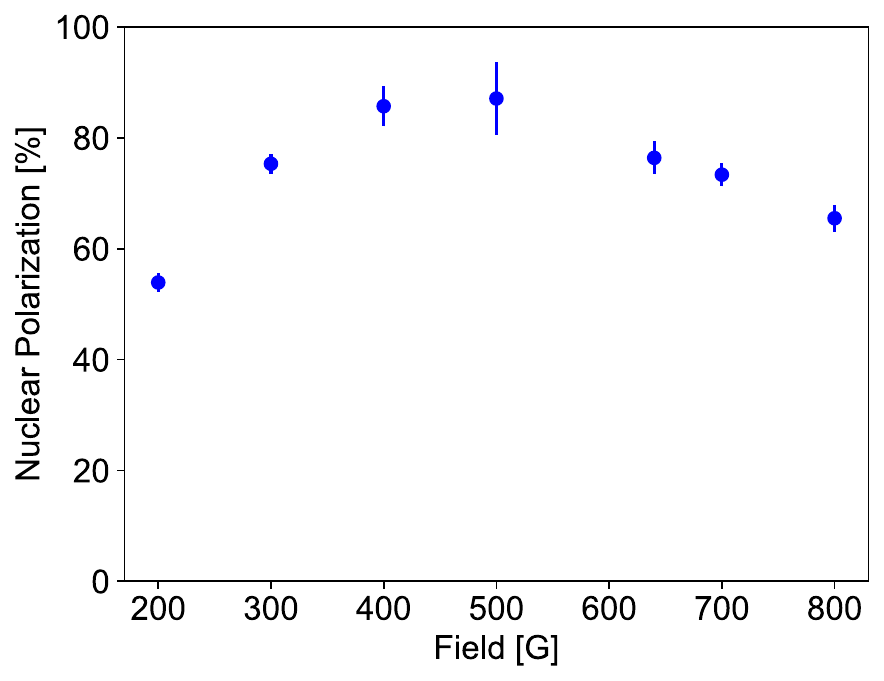}
    \caption{Measured nuclear polarization into $m_I = +1$ from 200-800~G. The polarization is calculated according to Eq.~F1 in the text, where 0\% corresponds to an unpolarized ensemble and 100\% corresponds to a perfectly polarized ensemble.}
\end{figure}
In our measurements, we take advantage of optical pumping near the ESLAC to polarize the nuclear spin to $m_I=+1$. While this is a simple and robust method of polarizing the nuclear spin, the polarization decreases for fields farther away from the ESLAC \cite{Fischer2013, Fischer2013a, Jacques2009}. In addition, polarization will also depend on the excitation power and initialization time. Imperfect polarization of the nuclear spin contributes to various features in our data, such as the loss in CNOT visibility at low and high fields described in the main text. 

To get a qualitative understanding of how this impacts our results, we measure the nuclear spin populations by fitting pulsed ODMR measurements of the electron spin at each field to a sum of three Lorentzians, constraining the hyperfine splitting between the transitions. The intensity of each of the three transitions ($I_0$, $I_{+1}$, and $I_{-1}$) is calculated by integrating the corresponding Lorentzians. Because we are calculating the polarization into a single state of a three level system, we define the polarization as
\begin{equation}
    P = 1- \frac{3}{2}\left(\frac{I_0+I_{-1}}{I_0+I_{-1}+I_{+1}}\right)
\end{equation}
where $P=1$ corresponds to perfect polarization in $m_I=+1$ and $P=0$ corresponds to the nuclear spin being completely unpolarized. 

We show a plot of the nuclear spin polarization for the fields that are used in our measurements in Figure F1. Far away from the ESLAC (at 200 and 800 G), the polarization is approximately 50\%. The maximum polarization is obtained near the ESLAC at around 85\%.

The amount of nuclear polarization that is given by a rate model (e.g. Ref \cite{Duarte2021, Manson2006, Robledo2011, Tetienne2012, Gupta2016}) has a strong dependence on the branching ratio in the ES ($\Gamma_{2,3}/\Gamma_4$) and the GS ($\Gamma_7/\Gamma_{5,6}$) at fields fields away from the ESLAC where the ES spin flip-flops are very inefficient. This does not affect the qualitative features of the trends we present (e.g. Figure A2), but makes direct quantitative comparisons between simulation and data difficult due to the discrepancy in the amount of nuclear polarization present at these fields (Lindblad (200 G) -- $\sim$12\%, Exp (200 G) -- $\sim$50\%). As a result, we restrict our quantitative comparison (Figure 6) between our Lindblad simulation and data to fields close to the ESLAC where there is good agreement in the amount nuclear polarization between our simulation and measurements (Lindblad (400 G) -- $\sim$77\%, Exp (400 G) -- 86\%).

\section{NV center ensemble $T_2^*$ Coherence}

\begin{figure}
    \centering
    \includegraphics[width=1\linewidth]{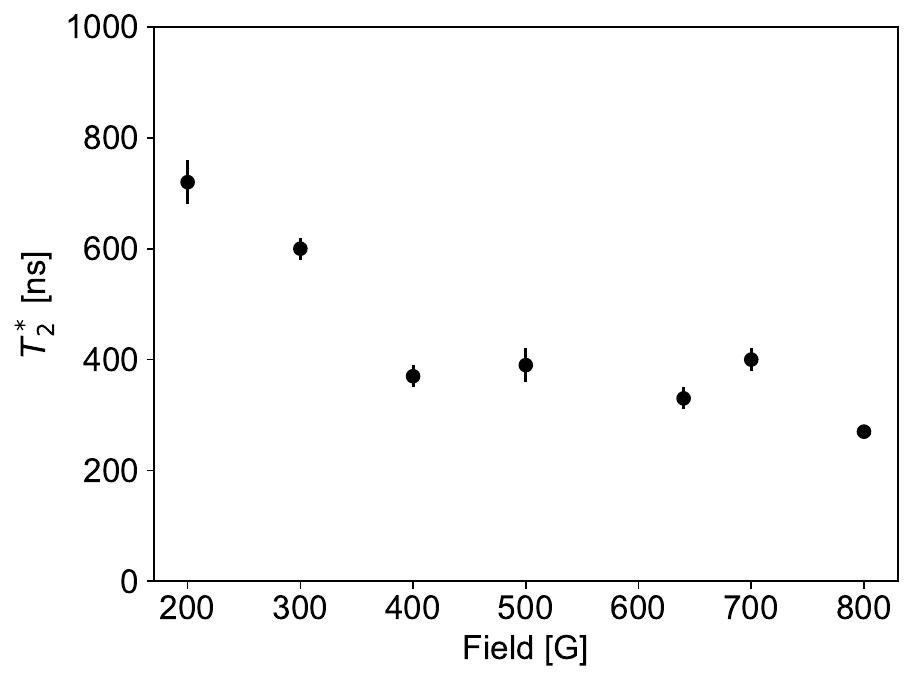}
    \caption{Field dependence of NV center electronic spin $T_2^*$. The maximum coherence is measured at 200~G is approximately 700~ns and decreases to approximately 400~ns at the ESLAC.}
\end{figure}

As discussed in the main text, a key parameter that limits the visibility of the CNOT readout is the inhomogeneous dephasing time $T_2^*$ of the electronic spin which limits the selectivity of our CNOT pulse with respect to the hyperfine split nuclear spin states. In general, $T_2^*$ is field dependent as it strongly depends on the coupling of the NV centers to the paramagnetic spin bath as well as the identity of the spins in the bath. For example, P1 centers near ESLAC are known to cause decreased coherence of NV centers \cite{Lazda2021}. 

To give an idea of how much the coherence time varies, we present the measured $T_2^*$ from 200-800~G in Figure G1. For our measurements, we fit the Ramsey measurements of the electron spin coherence to a sum of two exponentially decaying sinusoids:
\begin{align}
     C(t) = e^{\frac{t}{T_2^*}}[&A_1 \cos{(\omega t + \phi_1)}+ \\&A_2 \cos{((\omega t + C_\parallel) t + \phi_2})] + B   
\end{align}
where $\omega$ is the Ramsey oscillation frequency, $A_1$ and $A_2$ are the amplitudes of the sinusoids, $\phi_1$ and $\phi_2$ are phases that originate from factors such as imperfect pulse fidelities, and $B$ is an offset related to the spin populations in a decohered state and sources of background. The Ramsey oscillation frequencies of the two sinusoids are constrained by the GS hyperfine splitting $C_\parallel$. Physically, the two sinusoids represent the occupation in the two most populated nuclear hyperfine states after optical polarization ($\ket{0,+1}$ and $\ket{0,0}$). In this field range, the occupation in $\ket{0,-1}$ is negligible and is omitted from the fit. We find that our ensemble has the largest $T_2^*$ away from the ESLAC at 200-300~G (approximately 700~ns) and generally has a $T_2^*$ of approximately 400~ns near the ESLAC. We want to stress that while there is some variation in the coherence time across 200-800~G, we normalize for these variations by comparing our thermal state readout protocol with an analogous protocol with a $m_S = 0$ pure state as described in the main text (Figure 4).
\nocite{*}
\bibliographystyle{apsrev4-1}
\bibliographystyle{unsrt}
\bibliography{kuan_PRA_bib}

\begin{thebibliography}{10}

\bibitem{Barry2020}
John~F. Barry, Jennifer~M. Schloss, Erik Bauch, Matthew~J. Turner, Connor~A. Hart, Linh~M. Pham, and Ronald~L. Walsworth.
\newblock Sensitivity optimization for nv-diamond magnetometry.
\newblock {\em Reviews of Modern Physics}, 92(1):015004, March 2020.

\bibitem{Schloss2018}
Jennifer~M. Schloss, John~F. Barry, Matthew~J. Turner, and Ronald~L. Walsworth.
\newblock Simultaneous broadband vector magnetometry using solid-state spins.
\newblock {\em Physical Review Applied}, 10(3):034044, September 2018.

\bibitem{Taylor2008}
J.~M. Taylor, P.~Cappellaro, L.~Childress, L.~Jiang, D.~Budker, P.~R. Hemmer, A.~Yacoby, R.~Walsworth, and M.~D. Lukin.
\newblock High-sensitivity diamond magnetometer with nanoscale resolution.
\newblock {\em Nature Physics}, 4(10):810--816, September 2008.

\bibitem{Maze2008}
J.~R. Maze, P.~L. Stanwix, J.~S. Hodges, S.~Hong, J.~M. Taylor, P.~Cappellaro, L.~Jiang, M.~V.~Gurudev Dutt, E.~Togan, A.~S. Zibrov, A.~Yacoby, R.~L. Walsworth, and M.~D. Lukin.
\newblock Nanoscale magnetic sensing with an individual electronic spin in diamond.
\newblock {\em Nature}, 455(7213):644--647, October 2008.

\bibitem{Clevenson2015}
Hannah Clevenson, Matthew~E. Trusheim, Carson Teale, Tim Schröder, Danielle Braje, and Dirk Englund.
\newblock Broadband magnetometry and temperature sensing with a light-trapping diamond waveguide.
\newblock {\em Nature Physics}, 11(5):393--397, April 2015.

\bibitem{Mamin2013}
H.~J. Mamin, M.~Kim, M.~H. Sherwood, C.~T. Rettner, K.~Ohno, D.~D. Awschalom, and D.~Rugar.
\newblock Nanoscale nuclear magnetic resonance with a nitrogen-vacancy spin sensor.
\newblock {\em Science}, 339(6119):557--560, February 2013.

\bibitem{Dolde2011}
F.~Dolde, H.~Fedder, M.~W. Doherty, T.~Nöbauer, F.~Rempp, G.~Balasubramanian, T.~Wolf, F.~Reinhard, L.~C.~L. Hollenberg, F.~Jelezko, and J.~Wrachtrup.
\newblock Electric-field sensing using single diamond spins.
\newblock {\em Nature Physics}, 7(6):459--463, April 2011.

\bibitem{Michl2019}
Julia Michl, Jakob Steiner, Andrej Denisenko, André Bülau, André Zimmermann, Kazuo Nakamura, Hitoshi Sumiya, Shinobu Onoda, Philipp Neumann, Junichi Isoya, and Jörg Wrachtrup.
\newblock Robust and accurate electric field sensing with solid state spin ensembles.
\newblock {\em Nano Letters}, 19(8):4904--4910, July 2019.

\bibitem{Barson2021}
Michael S.~J. Barson, Lachlan~M. Oberg, Liam~P. McGuinness, Andrej Denisenko, Neil~B. Manson, Jörg Wrachtrup, and Marcus~W. Doherty.
\newblock Nanoscale vector electric field imaging using a single electron spin.
\newblock {\em Nano Letters}, 21(7):2962--2967, March 2021.

\bibitem{Toyli2012}
D.~M. Toyli, D.~J. Christle, A.~Alkauskas, B.~B. Buckley, C.~G. Van~de Walle, and D.~D. Awschalom.
\newblock Measurement and control of single nitrogen-vacancy center spins above 600 k.
\newblock {\em Physical Review X}, 2(3):031001, July 2012.

\bibitem{Toyli2013}
David~M. Toyli, Charles~F. de~las Casas, David~J. Christle, Viatcheslav~V. Dobrovitski, and David~D. Awschalom.
\newblock Fluorescence thermometry enhanced by the quantum coherence of single spins in diamond.
\newblock {\em Proceedings of the National Academy of Sciences}, 110(21):8417--8421, May 2013.

\bibitem{Fukami2019}
M.~Fukami, C.G. Yale, P.~Andrich, X.~Liu, F.J. Heremans, P.F. Nealey, and D.D. Awschalom.
\newblock All-optical cryogenic thermometry based on nitrogen-vacancy centers in nanodiamonds.
\newblock {\em Physical Review Applied}, 12(1):014042, July 2019.

\bibitem{Wood2018}
Alexander~A. Wood, Emmanuel Lilette, Yaakov~Y. Fein, Nikolas Tomek, Liam~P. McGuinness, Lloyd C.~L. Hollenberg, Robert~E. Scholten, and Andy~M. Martin.
\newblock Quantum measurement of a rapidly rotating spin qubit in diamond.
\newblock {\em Science Advances}, 4(5), May 2018.

\bibitem{Soshenko2021}
Vladimir~V. Soshenko, Stepan~V. Bolshedvorskii, Olga Rubinas, Vadim~N. Sorokin, Andrey~N. Smolyaninov, Vadim~V. Vorobyov, and Alexey~V. Akimov.
\newblock Nuclear spin gyroscope based on the nitrogen vacancy center in diamond.
\newblock {\em Physical Review Letters}, 126(19):197702, May 2021.

\bibitem{Jarmola2021}
Andrey Jarmola, Sean Lourette, Victor~M. Acosta, A.~Glen Birdwell, Peter Blümler, Dmitry Budker, Tony Ivanov, and Vladimir~S. Malinovsky.
\newblock Demonstration of diamond nuclear spin gyroscope.
\newblock {\em Science Advances}, 7(43), October 2021.

\bibitem{Ladd2005}
T.~D. Ladd, D.~Maryenko, Y.~Yamamoto, E.~Abe, and K.~M. Itoh.
\newblock Coherence time of decoupled nuclear spins in silicon.
\newblock {\em Physical Review B}, 71(1):014401, January 2005.

\bibitem{Park2017}
Jee~Woo Park, Zoe~Z. Yan, Huanqian Loh, Sebastian~A. Will, and Martin~W. Zwierlein.
\newblock Second-scale nuclear spin coherence time of ultracold 23 na 40 k molecules.
\newblock {\em Science}, 357(6349):372--375, July 2017.

\bibitem{Shaham2022}
R.~Shaham, O.~Katz, and O.~Firstenberg.
\newblock Strong coupling of alkali-metal spins to noble-gas spins with an hour-long coherence time.
\newblock {\em Nature Physics}, 18(5):506--510, April 2022.

\bibitem{Maurer2012}
P.~C. Maurer, G.~Kucsko, C.~Latta, L.~Jiang, N.~Y. Yao, S.~D. Bennett, F.~Pastawski, D.~Hunger, N.~Chisholm, M.~Markham, D.~J. Twitchen, J.~I. Cirac, and M.~D. Lukin.
\newblock Room-temperature quantum bit memory exceeding one second.
\newblock {\em Science}, 336(6086):1283--1286, June 2012.

\bibitem{Steger2012}
M.~Steger, K.~Saeedi, M.~L.~W. Thewalt, J.~J.~L. Morton, H.~Riemann, N.~V. Abrosimov, P.~Becker, and H.-J. Pohl.
\newblock Quantum information storage for over 180 s using donor spins in a 28 si “semiconductor vacuum”.
\newblock {\em Science}, 336(6086):1280--1283, June 2012.

\bibitem{Jarmola2012}
A.~Jarmola, V.~M. Acosta, K.~Jensen, S.~Chemerisov, and D.~Budker.
\newblock Temperature- and magnetic-field-dependent longitudinal spin relaxation in nitrogen-vacancy ensembles in diamond.
\newblock {\em Physical Review Letters}, 108(19):197601, May 2012.

\bibitem{Wang2023}
Guoqing Wang, Ariel~Rebekah Barr, Hao Tang, Mo~Chen, Changhao Li, Haowei Xu, Andrew Stasiuk, Ju~Li, and Paola Cappellaro.
\newblock Characterizing temperature and strain variations with qubit ensembles for their robust coherence protection.
\newblock {\em Physical Review Letters}, 131(4):043602, July 2023.

\bibitem{Wang2024}
Guoqing Wang, Minh-Thi Nguyen, Dane~W. deQuilettes, Eden Price, Zhiyao Hu, Danielle~A. Braje, and Paola Cappellaro.
\newblock Emulated nuclear spin gyroscope with $^{15}$nv centers in diamond.
\newblock {\em arxiv}, January 2024.

\bibitem{Chen2018}
Mo~Chen, Won Kyu~Calvin Sun, Kasturi Saha, Jean-Christophe Jaskula, and Paola Cappellaro.
\newblock Protecting solid-state spins from a strongly coupled environment.
\newblock {\em New Journal of Physics}, 20(6):063011, June 2018.

\bibitem{Jiang2009}
L.~Jiang, J.~S. Hodges, J.~R. Maze, P.~Maurer, J.~M. Taylor, D.~G. Cory, P.~R. Hemmer, R.~L. Walsworth, A.~Yacoby, A.~S. Zibrov, and M.~D. Lukin.
\newblock Repetitive readout of a single electronic spin via quantum logic with nuclear spin ancillae.
\newblock {\em Science}, 326(5950):267--272, October 2009.

\bibitem{Neumann2010}
Philipp Neumann, Johannes Beck, Matthias Steiner, Florian Rempp, Helmut Fedder, Philip~R. Hemmer, Jörg Wrachtrup, and Fedor Jelezko.
\newblock Single-shot readout of a single nuclear spin.
\newblock {\em Science}, 329(5991):542--544, July 2010.

\bibitem{Poggiali2017}
F.~Poggiali, P.~Cappellaro, and N.~Fabbri.
\newblock Measurement of the excited-state transverse hyperfine coupling in nv centers via dynamic nuclear polarization.
\newblock {\em Physical Review B}, 95(19):195308, May 2017.

\bibitem{Doherty2013}
Marcus~W. Doherty, Neil~B. Manson, Paul Delaney, Fedor Jelezko, Jörg Wrachtrup, and Lloyd~C.L. Hollenberg.
\newblock The nitrogen-vacancy colour centre in diamond.
\newblock {\em Physics Reports}, 528(1):1--45, July 2013.

\bibitem{Duarte2021}
H.~Duarte, H.~T. Dinani, V.~Jacques, and J.~R. Maze.
\newblock Effect of intersystem crossing rates and optical illumination on the polarization of nuclear spins close to nitrogen-vacancy centers.
\newblock {\em Physical Review B}, 103(19):195443, May 2021.

\bibitem{Robledo2011}
Lucio Robledo, Hannes Bernien, Toeno van~der Sar, and Ronald Hanson.
\newblock Spin dynamics in the optical cycle of single nitrogen-vacancy centres in diamond.
\newblock {\em New Journal of Physics}, 13(2):025013, February 2011.

\bibitem{Gupta2016}
A.~Gupta, L.~Hacquebard, and L.~Childress.
\newblock Efficient signal processing for time-resolved fluorescence detection of nitrogen-vacancy spins in diamond.
\newblock {\em Journal of the Optical Society of America B}, 33(3):B28, January 2016.

\bibitem{Fischer2013}
Ran Fischer, Christian~O. Bretschneider, Paz London, Dmitry Budker, David Gershoni, and Lucio Frydman.
\newblock Bulk nuclear polarization enhanced at room temperature by optical pumping.
\newblock {\em Physical Review Letters}, 111(5):057601, July 2013.

\bibitem{Fischer2013a}
Ran Fischer, Andrey Jarmola, Pauli Kehayias, and Dmitry Budker.
\newblock Optical polarization of nuclear ensembles in diamond.
\newblock {\em Physical Review B}, 87(12):125207, March 2013.

\bibitem{Jacques2009}
V.~Jacques, P.~Neumann, J.~Beck, M.~Markham, D.~Twitchen, J.~Meijer, F.~Kaiser, G.~Balasubramanian, F.~Jelezko, and J.~Wrachtrup.
\newblock Dynamic polarization of single nuclear spins by optical pumping of nitrogen-vacancy color centers in diamond at room temperature.
\newblock {\em Physical Review Letters}, 102(5):057403, February 2009.

\bibitem{Steiner2010}
M.~Steiner, P.~Neumann, J.~Beck, F.~Jelezko, and J.~Wrachtrup.
\newblock Universal enhancement of the optical readout fidelity of single electron spins at nitrogen-vacancy centers in diamond.
\newblock {\em Physical Review B}, 81(3):035205, January 2010.

\bibitem{Jarmola2020}
A.~Jarmola, I.~Fescenko, V.~M. Acosta, M.~W. Doherty, F.~K. Fatemi, T.~Ivanov, D.~Budker, and V.~S. Malinovsky.
\newblock Robust optical readout and characterization of nuclear spin transitions in nitrogen-vacancy ensembles in diamond.
\newblock {\em Physical Review Research}, 2(2):023094, April 2020.

\bibitem{Thew2002}
R.~T. Thew, K.~Nemoto, A.~G. White, and W.~J. Munro.
\newblock Qudit quantum-state tomography.
\newblock {\em Physical Review A}, 66(1):012303, July 2002.

\bibitem{Hofmann2004}
Holger~F. Hofmann and Shigeki Takeuchi.
\newblock Quantum-state tomography for spin-l systems.
\newblock {\em Physical Review A}, 69(4):042108, April 2004.

\bibitem{Acosta2009}
V.~M. Acosta, E.~Bauch, M.~P. Ledbetter, C.~Santori, K.-M.~C. Fu, P.~E. Barclay, R.~G. Beausoleil, H.~Linget, J.~F. Roch, F.~Treussart, S.~Chemerisov, W.~Gawlik, and D.~Budker.
\newblock Diamonds with a high density of nitrogen-vacancy centers for magnetometry applications.
\newblock {\em Physical Review B}, 80(11):115202, September 2009.

\bibitem{Bauch2018}
Erik Bauch, Connor~A. Hart, Jennifer~M. Schloss, Matthew~J. Turner, John~F. Barry, Pauli Kehayias, Swati Singh, and Ronald~L. Walsworth.
\newblock Ultralong dephasing times in solid-state spin ensembles via quantum control.
\newblock {\em Physical Review X}, 8(3):031025, July 2018.

\bibitem{Bauch2020}
Erik Bauch, Swati Singh, Junghyun Lee, Connor~A. Hart, Jennifer~M. Schloss, Matthew~J. Turner, John~F. Barry, Linh~M. Pham, Nir Bar-Gill, Susanne~F. Yelin, and Ronald~L. Walsworth.
\newblock Decoherence of ensembles of nitrogen-vacancy centers in diamond.
\newblock {\em Physical Review B}, 102(13):134210, October 2020.

\bibitem{Nielsen2012}
Michael~A. Nielsen and Isaac~L. Chuang.
\newblock {\em Quantum Computation and Quantum Information: 10th Anniversary Edition}.
\newblock Cambridge University Press, June 2012.

\bibitem{Fuchs2012}
G.~D. Fuchs, A.~L. Falk, V.~V. Dobrovitski, and D.~D. Awschalom.
\newblock Spin coherence during optical excitation of a single nitrogen-vacancy center in diamond.
\newblock {\em Physical Review Letters}, 108(15):157602, April 2012.

\bibitem{Howard2006}
M~Howard, J~Twamley, C~Wittmann, T~Gaebel, F~Jelezko, and J~Wrachtrup.
\newblock Quantum process tomography and linblad estimation of a solid-state qubit.
\newblock {\em New Journal of Physics}, 8(3):33--33, March 2006.

\bibitem{O’Brien2004}
J.~L. O’Brien, G.~J. Pryde, A.~Gilchrist, D.~F.~V. James, N.~K. Langford, T.~C. Ralph, and A.~G. White.
\newblock Quantum process tomography of a controlled-not gate.
\newblock {\em Physical Review Letters}, 93(8):080502, August 2004.

\bibitem{Harris2023}
Isaac~B.W. Harris, Cathryn~P. Michaels, Kevin~C. Chen, Ryan~A. Parker, Michael Titze, Jesús Arjona~Martínez, Madison Sutula, Ian~R. Christen, Alexander~M. Stramma, William Roth, Carola~M. Purser, Martin~Hayhurst Appel, Chao Li, Matthew~E. Trusheim, Nicola~L. Palmer, Matthew~L. Markham, Edward~S. Bielejec, Mete Atatüre, and Dirk Englund.
\newblock Hyperfine spectroscopy of isotopically engineered group-iv color centers in diamond.
\newblock {\em PRX Quantum}, 4(4):040301, October 2023.

\bibitem{KuateDefo2021}
Rodrick Kuate~Defo, Efthimios Kaxiras, and Steven~L. Richardson.
\newblock Calculating the hyperfine tensors for group-iv impurity-vacancy centers in diamond using hybrid density functional theory.
\newblock {\em Physical Review B}, 104(7):075158, August 2021.

\bibitem{Soshenko2020}
Vladimir~V. Soshenko, Vadim~V. Vorobyov, Stepan~V. Bolshedvorskii, Olga Rubinas, Ivan Cojocaru, Boris Kudlatsky, Anton~I. Zeleneev, Vadim~N. Sorokin, Andrey~N. Smolyaninov, and Alexey~V. Akimov.
\newblock Temperature drift rate for nuclear terms of the nv-center ground-state hamiltonian.
\newblock {\em Physical Review B}, 102(12):125133, September 2020.

\bibitem{Lourette2023}
Sean Lourette, Andrey Jarmola, Victor~M. Acosta, A.~Glen Birdwell, Dmitry Budker, Marcus~W. Doherty, Tony Ivanov, and Vladimir~S. Malinovsky.
\newblock Temperature sensitivity of 14n - v and 15n - v ground-state manifolds.
\newblock {\em Physical Review Applied}, 19(6):064084, June 2023.

\bibitem{Breuer2007}
Heinz-Peter Breuer and Francesco Petruccione.
\newblock {\em The Theory of Open Quantum Systems}.
\newblock Oxford University Press, January 2007.

\bibitem{Qutip}
J.~R. Johansson, P.~D. Nation, and F.~Nori.
\newblock Qutip 2: A python framework for the dynamics of open quantum systems.
\newblock {\em Computer Physics Communications}, 184:1234, 2013.

\bibitem{McCullian2022}
B.A. McCullian, H.F.H. Cheung, H.Y. Chen, and G.D. Fuchs.
\newblock Quantifying the spectral diffusion of n- v centers by symmetry.
\newblock {\em Physical Review Applied}, 18(6):064011, December 2022.

\bibitem{Manson2006}
N.~B. Manson, J.~P. Harrison, and M.~J. Sellars.
\newblock Nitrogen-vacancy center in diamond: Model of the electronic structure and associated dynamics.
\newblock {\em Physical Review B}, 74(10):104303, September 2006.

\bibitem{Tetienne2012}
J.~Tetienne, L.~Rondin, P.~Spinicelli, M.~Chipaux, T.~Debuisschert, J.~Roch, and V.~Jacques.
\newblock Magnetic-field-dependent photodynamics of single {NV} defects in diamond: an application to qualitative all-optical magnetic imaging.
\newblock {\em New Journal of Physics}, 14(10):103033, October 2012.

\bibitem{Lazda2021}
Reinis Lazda, Laima Busaite, Andris Berzins, Janis Smits, Florian Gahbauer, Marcis Auzinsh, Dmitry Budker, and Ruvin Ferber.
\newblock Cross-relaxation studies with optically detected magnetic resonances in nitrogen-vacancy centers in diamond in external magnetic field.
\newblock {\em Physical Review B}, 103(13):134104, April 2021.

\end{thebibliography}
\end{document}